\documentclass[12pt]{iopart}

\usepackage{iopams}  
\expandafter\let\csname equation*\endcsname\relax
\expandafter\let\csname endequation*\endcsname\relax

\usepackage[linesnumbered,ruled,vlined]{algorithm2e}
\usepackage{amsmath}
\usepackage{amssymb} 
\usepackage{amsthm}
\usepackage{bm,mathrsfs,bbm} 

\usepackage[utf8]{inputenc}  
\usepackage[T1]{fontenc}     
\usepackage[british]{babel}  
\usepackage[sc,osf]{mathpazo}
\usepackage[scaled=0.86]{berasans}  
\usepackage[colorlinks=true, allcolors=blue, urlcolor=blue]{hyperref}  
\usepackage{graphicx} 
\usepackage[babel]{microtype}  

\usepackage{xspace}  
\usepackage{pgfplots}
\usepackage{xcolor,colortbl}
\usepackage{array}
\usepackage{bigstrut}
\usepackage{cases}
\usepackage{graphicx}
\usepackage{caption}
\usepackage{subcaption}
\usepackage{ragged2e}
\DeclareCaptionJustification{justified}{\justifying}
\usepackage{tabularx}
\newcolumntype{C}{>{\centering\arraybackslash}X}

\usepackage{lipsum}

\newcommand{\be}{\begin{equation}}
\newcommand{\ee}{\end{equation}}
\newcommand{\bea}{\begin{eqnarray}}
\newcommand{\eea}{\end{eqnarray}}
\newcommand{\bes}{\begin{equation*}}
\newcommand{\ees}{\end{equation*}}
\newcommand{\beas}{\begin{eqnarray*}}
\newcommand{\eeas}{\end{eqnarray*}}
\newcommand{\bpm}{\begin{pmatrix}}
\newcommand{\epm}{\end{pmatrix}}

\usepackage{verbatim}
\usepackage{tcolorbox}
\usepackage{listings}

\makeatletter
\newcommand{\ostar}{\mathbin{\mathpalette\make@circled\star}}
\newcommand{\make@circled}[2]{%
  \ooalign{$\m@th#1\smallbigcirc{#1}$\cr\hidewidth$\m@th#1#2$\hidewidth\cr}%
}
\newcommand{\smallbigcirc}[1]{%
  \vcenter{\hbox{\scalebox{0.77778}{$\m@th#1\bigcirc$}}}%
}
\makeatother

\newtheorem{thm}{Theorem}
\newtheorem{lem}{Lemma}

\newtheorem{defn}{Definition}
\newtheorem{pr}{Property}

\begin{document}

\title{Quantum Advantages in $\mathbf{(n,d)\!\mapsto\!1}$ Random Access Codes}

\author{Andris Ambainis \& Dmitry Kravchenko}
\ead{andris.ambainis@lu.lv \& kravchenko@gmail.com}
\address{Faculty of Computing, University of Latvia, Raina bulv. 19, Riga, LV-1586, Latvia}

\author{Sk Sazim}
\ead{ssazim@cft.edu.pl}
\address{Center for Theoretical Physics, Polish Academy of Sciences, Aleja Lotnik\'{o}w 32/46, 02-668 Warsaw, Poland}

\author{ Joonwoo Bae \& Ashutosh Rai}
\ead{joonwoo.bae@kaist.ac.kr \& ashutosh.rai@kaist.ac.kr}
\address{School of Electrical Engineering, Korea Advanced Institute of Science and Technology (KAIST), $291$ Daehak-ro, Yuseong-gu, Daejeon $34141$, Republic of Korea }


\begin{abstract}
A random access code (RAC), corresponding to a communication primitive with various applications in quantum information theory, is an instance of a preparation-and-measurement scenario. In this work, we consider $(n,d)$-RACs constituting an $n$-length string, constructed from a $d$ size set of letters, and send an encoding of the string in a single $d$-level physical system and present their quantum advantages. We first characterize optimal classical RACs and prove that a known classical strategy, called \emph{majority-encoding-identity-decoding}, is optimal. We then construct a quantum protocol by exploiting only two incompatible measurements (the minimal requirement) and show the advantages beyond the classical one. We also discuss the generality of our results and whether quantum advantages are valid for all types of $(n, d) \!\mapsto\! 1$ RACs.
\end{abstract}

%
%
%
\maketitle
%
%



\section{Introduction} 

Finding information processing tasks for which quantum resources provide an advantage over their classical counterparts is an underlying theme for a wide range of research problems in quantum information science. These problems, to mention a few, cover topics as diverse as designing efficient quantum algorithms \cite{qalgo0,qalgo01, qalgo1,qalgo2,qalgo3}, advancing quantum cryptography \cite{qcrypt0,qcrypt01,qcrypt1,qcrypt2,qcrypt3,Feix-Brukner1}, improving communication complexity using quantum resources \cite{comu1,comu2,comu3,Gupta-Deba-Cabello}, quantum state discrimination~\cite{Junu-Hwang-Han-PRL2011,Bae_2013,Bae_2015}, and numerous other problems in quantum computing and quantum information \cite{Nielson,Doriguello2021quantumrandomaccess,RK-MB,Piveteau+2022,Ambainis+2019}.

 A random access code (RAC), a communication primitive, is one of the first instances where quantum advantages are evident in information theory. In particular, an $(n,d) \overset{p}{\mapsto} 1$ RAC involves two parties: a sender, referred to as Alice who receives a random word of length $n$, where each letter in the word is an element of a  set with cardinality $d$, and encodes her word in a single $d$-dimensional physical system. A receiver, called Bob, aims to learn the value of a randomly selected letter from the word with a probability of at least $p$. A RAC protocol can be quantified by the worst-case success probability $p$.

A quantum generalization of random access codes, known as quantum random access codes (QRACs) \cite{Wiesner1983,Nayak1999,Ambainis1999,Ambainis2002,Hayashi2006,Tavakoli}, was first introduced as conjugate coding \cite{Wiesner1983} and utilizes quantum systems shared between two parties. The concept was formally explored in Ref. \cite{Ambainis2002}, where it was demonstrated that QRACs offer quantum advantages, although they are subject to a natural limitation imposed by the Holevo theorem.

In the initial formulation of a QRAC, the figure of merit was the worst-case guessing probability while a sender and a receiver are not allowed to have access to shared randomness \cite{Ambainis1999,Nayak1999,Ambainis2002,Hayashi2006}. Under the assumptions, quantum advantages were shown for $(2,2)\overset{p}{\mapsto}1$ RACs~\cite{Ambainis1999,Ambainis2002} and $(3,2)\overset{p}{\mapsto}1$ RACs~\cite{Hayashi2006}. Then, a no-go theorem has been presented: a $(n,2)\overset{p}{\mapsto}1$ QRAC for $n\geq 4$ does not have an advantage over its classical counterpart in terms of the worst-case probability \cite{Hayashi2006}.

In Ref. \cite{Ambainis2008}, it has been shown that for  $(n,2)\overset{p}{\mapsto}1$ RACs, the maximum value of worst-case success probability $p$ while two parties are allowed to have access to shared randomness is equal to the maximum value of an average success probability. Since then, a QRAC has been formulated such that two parties have access to shared randomness and the figure of merit is chosen as an average success probability. With the formulation, a RAC can be denoted by $(n,d)\!\mapsto\! 1$, throughout. Along the line, much has been devoted to characterizing quantum advantages, see e.g., \cite{Tavakoli} and \cite{FarkasMiklinTavakoli2024}.  Various applications in quantum information theory find the usefulness of the results in RACs, including quantum automata \cite{Ambainis1999,Nayak1999} network coding \cite{app2}, semi-device independent QKD \cite{app4,app5} and foundations of quantum mechanics \cite{app6,app7,Tavakoli,Heinosaari_2022, Deba-Horo1,Bera-Maju1,Mancinska2022,Sazim1}.  

Let us reiterate the motivation of a QRAC from the beginning \cite{Ambainis1999, Ambainis2002} that has questioned quantum advantages in a two-party communication protocol while the Holevo theorem asserts quantum advantages are unlikely in a point-to-point communication scenario. To this end, firstly, a QRAC protocol has been devised such that while classical information is encoded in a quantum state, the receiver's goal is to extract only a selective (partial) information by choosing to apply a decoding measurement from a set of incompatible measurements. Then, a quantum advantage has been proven with rigour by first proving the classical optimality and then showing that a quantum strategy beats the classical limit. Since the first results in QRAC protocols \cite{Ambainis1999, Ambainis2002}, quantum advantages of $(n,d)\mapsto 1$ QRAC protocols in general with rigour may be yet far from reaching due to the lack of a full characterization of classical optimal strategies. In fact, little is known about how to characterize the all possible classical optimal protocols, which may provides  insights on how to construct QRACs protocols. All these make it a key problem to show quantum advantages in a QRAC protocol with rigour. 

In this work, we show that $(n,d)\!\mapsto\! 1$ RACs contain quantum advantages in general by characterizing the limitations of classical RACs and considering quantum strategies. The characterization includes a proof to the conjecture, addressed in Ref. \cite{Tavakoli}, that a classical strategy, called \emph{majority-encoding-identity-decoding} is optimal. We show that two incompatible measurements of a receiver is possibly suffice to attain quantum advantages over classical optimal strategies. We also investigate quantum strategies by comparing recent results in Ref. \cite{FarkasMiklinTavakoli2024}.

This work is structured as follows. In Sec. \ref{sec:2}, we summarize the scenario of $(n,d)\!\mapsto\! 1$ RACs. In Sec.~\ref{sec:3}, we characterize optimal classical strategies and show that a known strategy, called majority-encoding and identity-decoding, is optimal. Also, we provide an algorithm to compute the values of optimal average success in classical protocols. In Sec.~\ref{sec:4}, we provide a quantum protocol with the decoding with two incompatible measurements and show that the quantum protocol beats classical strategies. In Sec.~\ref{sec:5}, we compare the results with recent results. We then summarize our results in Sec. \ref{sec:6}.

\section{ Random Access Codes}
\label{sec:2}
In this section, let us summarize the scenario of RACs. We consider RAC protocols by assuming that two parties can access \emph{shared randomness}. A sender, Alice, prepares a message of length $n$ uniformly at random, 
\bea
x = x_1x_2 \ldots x_n,~~\mathrm{where}~~x_i\in X=\{0,1,\ldots,d\!\!-\!\!1\}.\label{eq:msg}
\eea 
The other party, Bob, selects an element
\bea
j\in\mathcal{J}=\{1,2,\ldots,n\} 
\eea
randomly and uniformly. With a selection of $j$, Bob attempts to guess the $j$-th letter, $x_j$, in the message in Eq. (\ref{eq:msg}). To this end, two parties are supposed to deliver a single system in one direction from Alice to Bob, where the system is $d$-dimensional and either classical or quantum. The question lies in comparing classical and quantum resources in an RAC protocol. The scenario is depicted in Fig. \ref{fig1}.

Two parties have control over two tasks, called their strategy, encoding and decoding. 
	\begin{itemize}
 \vspace{0.2cm}
			\item \emph{Encoding}:~Alice sends a system prepared by an encoding $\mathcal{E}$, i.e, $\mathcal{E}(x)$ for a word $x\in X^n$, where the system is $d$-dimensional.
  A QRAC protocol consists of an encoding $\mathcal{E}_Q$ that corresponds to quantum state preparation which maps a word $x$ to a $d$-dimensional quantum state $\mathcal{E}_Q(x)\in \mathbb{C}^d$. On the other hand, an encoding $\mathcal{E}_{cl}$ in a classical RAC protocol is a probabilistic mapping of words $x$ to the alphabet set $X=\{0,\cdots,d\!-\!1\}$ according to some probability distribution $\{p_x(0),\cdots,p_x(d\!-\!1)\}$.
    \vspace{0.2cm}
		\item \emph{Decoding}:~ Bob can manipulate a system sent by Alice to give an answer that makes a guess about a letter for his selection $j\in\mathcal{J}$. His local operation is signified by a decoding $\mathcal{D}(\mathcal{E}(x),j)$. In a QRAC protocol a decoding is realized by a set of $d$-outcome measurements $\{\mathcal{D}_Q(j):~j\in\{1,\cdots,n\}\}$ on the quantum state $\mathcal{E}_Q(x)$. For guessing the $j$-th letter in $x$, Bob applies the measurement $\mathcal{D}_Q(j)\equiv \{\Pi_0^{j},\Pi_1^{j},\cdots,\Pi_{d\!-\!1}^{j}\}$ where $\sum^{d\!-\!1}_{\kappa=0}\Pi_{\kappa}^{j}=\mathbbm{1}$, and $\Pi_{\kappa}^{j}$ are positive-operator-valued-measure (POVM) elements on a $d$-dimensional state space, and by applying Born's rule for quantum probabilities answers $g\in\{0,\cdots,d\!-\!1\}$ as his guess for $x_j$ when the effect $\Pi_{g}^{j}$ clicks. In classical RAC, a decoding $\mathcal{D}_{cl}$ is a probabilstic mapping of an ordered pair $z=(\mathcal{E}_{cl}(x),j)$ to the alphabet set $X=\{0,\cdots,d\!-\!1\}$ according to some probability distribution $\{p_z(0),\cdots,p_z(d\!-\!1)\}$.
	\end{itemize}
 
	\begin{figure}[t!]
		\begin{center}
			\includegraphics[scale=0.25]{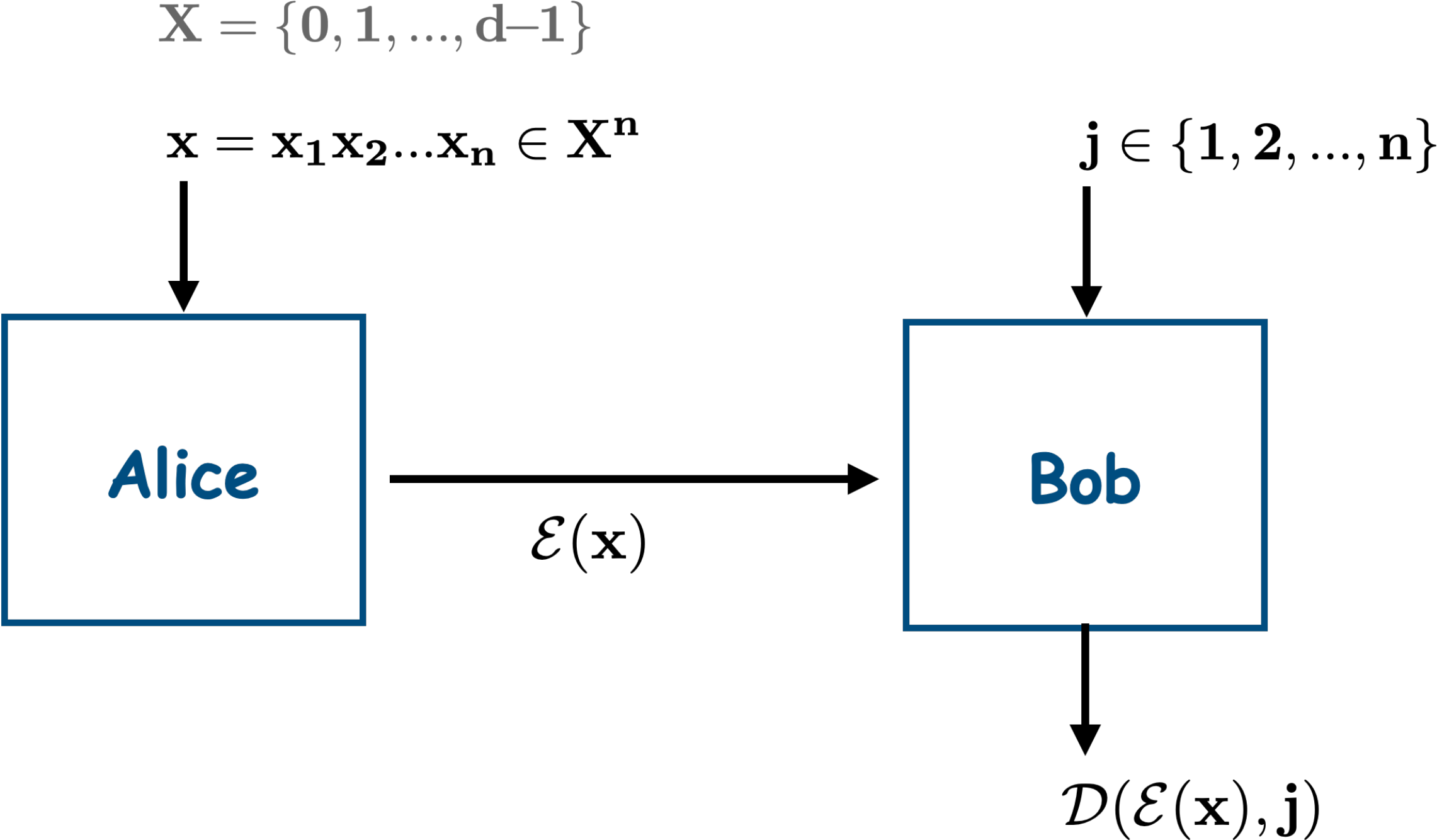}
			\caption{{\small In a $(n,d)\!\mapsto\! 1$ RAC: Alice receives a uniformly-random $n$-letter word $x=x_1x_2...x_n \in X^n$ where $X=\{0,1,...,d\!\!-\!\!1\}$. Bob receives uniformly at random an index $j\in \{1,2,...,n\}$ and his task is to guess the letter $x_j$ in Alice's word. Alice can encode her word and communicate to Bob only a single $d$-dimentional quantum or classical system $\mathcal{E}(x)$. The success of the task is measured by the average probability of a correct guess. A joint strategy for the task consists of an encoding $\mathcal{E}(x)$ by Alice and a decoding $\mathcal{D}(\mathcal{E}(x),j)$ by Bob.} }
			\label{fig1}
		\end{center}
	\end{figure}
 

Let $\mathcal{S}$ denote a joint strategy (a protocol) by a pair, an encoding and a decoding so that it gives a conclusive guess at the end of a round of the protocol. Then, the protocol $\mathcal{S}$ is written by an ordered pair
\bea
\mathcal{S}\equiv (\mathcal{E}, \mathcal{D}),\mbox{~~where~}
S(x,j) = \mathcal{D}(\mathcal{E}(x),j) 
\eea
for a word $x \in X^n$ and a selection $j\in \mathcal{J}$. Note that in general both encoding and decoding are described by probabilistic maps, i.e., $\mathcal{S}$ in general is a random variable and $\mathcal{S}(x,j)$ results in a outcome in the set $\{0,\cdots,d\!-\!1\}$ following some probability distribution.

RAC protocols are quantified by a value of strategy $\mathcal{S}$, which is an average probability of making a correct guess over all possible inputs: 	\bea\label{value-rac}
		\mbox{Value}(\mathcal{S})  &\equiv&\overline{\mathrm{P}}_{\mathcal{S}} ~~\mathrm{where} \nonumber\\
\overline{\mathrm{P}}_{\mathcal{S}}   &=& \frac{1}{nd^n}\sum_{x \in X^n}\sum_{j=1}^n \mathrm{P}[\mathcal{S}(x,j)=x_j]  \label{eq:gue}
\eea
and $\mathrm{P}[\mathcal{S}(x,j)=x_j]$ is the probability that a guess of the strategy $S$ is correct for a message $x$ of Alice and a selection $j$ of Bob.

	\section{ Optimal classical strategies in $\mathbf{(n,d)\!\mapsto\! 1}$ random access codes}
\label{sec:3}

Knowing the optimal classical strategies is of fundamental importance since it provides the benchmark for showing quantum advantage. In this section, we characterize all possible optimal classical strategies for the $(n,d)\!\mapsto\! 1$ RACs.

\subsection{Optimal classical strategies}
 An optimal strategy could be either randomized or deterministic. In a deterministic strategy, first Alice encodes her words $x = x_1 x_2 \ldots x_n \in X^n$ by applying some deterministic function
	\bea
	\mathcal{E}:X^n\!\rightarrow \! X 
	\eea
	and sends to Bob $y=\mathcal{E}(x)\in X$. Then, Bob decodes a received message, which is realized by a deterministic function 
	\bea
	\mathcal{D}:(X,\mathcal{J})\!\rightarrow\! X, 
	\eea
	i.e., Bob's guess for the $j$-th letter in Alice's word is given by $\mathcal{D}\left(y,j\right) \in X$. The set of all possible deterministic encodings and decodings can be written as respectively,  
	\bea
	\mathbb{E}=\{\mathcal{E}^u: u\in {\rm U}=\{1,\cdots,d^{d^n}\}\},\\
	\mathbb{D}= \{\mathcal{D}^v: v\in{\rm V}=\{1,\cdots,d^{nd}\}\}.
	\eea
	Then, the set of all possible deterministic protocols are
	\bea
	\mathbb{S}=\{\mathcal{S}^{uv}\equiv (\mathcal{E}^u,\mathcal{D}^v ):~u\in{\rm U}~\mbox{and}~ v\in{\rm V}\},
	\eea
	where $\mathcal{S}^{uv}(x,j)=\mathcal{D}^v(\mathcal{E}^u(x),j)$. Suppose Alice and Bob has access to a shared random variable $\Lambda=\{\lambda_{uv}: ~u\in{\rm U}~\mbox{and}~ v\in{\rm V}\}$ such that $\lambda_{uv}$ is shared between Alice and Bob with a probability $p_{uv}(\lambda_{uv})$, then sender and receivers, by using $\Lambda$, can correlate the deterministic protocols and implement a randomized protocol $\mathcal{S}=\{\{\mathcal{S}^{uv}, p_{uv}\}:~u\in{\rm U}~\mbox{and}~ v\in{\rm V}\}$. Thus, any randomized strategy can be realized as a probabilistic mixture of deterministic strategies by Alice and Bob utilizing shared randomness. The average success probability in Eq.~(\ref{eq:gue}) for a randomized protocol is
	\bea
	\overline{{\rm P}}_{\mathcal{S}}:=\sum_{u,v}p_{uv}~\overline{{\rm P}}_{\mathcal{S}^{uv}}\leq \max_{S^{uv} }\overline{{\rm P}}_{\mathcal{S}^{uv}},
	\eea
	which is upper bounded by the maximum average success probability over all the deterministic protocols. Therefore, it is sufficient to maximize classical average success probability over the set of all the deterministic protocols. So from here we consider only deterministic classical strategies. Then, for a message $y$ sent from Alice, Bob must have a $n$-length string 
\begin{equation}   
	 \mathcal{D}_y=\mathcal{D}\left(y,1\right)~\mathcal{D}\left(y,2\right)~ \ldots ~\mathcal{D}\left(y,n\right). 
\end{equation}
The decoding strategy of Bob is completely described by a $d\times n$ matrix $[\mathcal{D}(y,j)]$. Upon receiving a message $y$ from Alice, Bob looks into the $y$-th row of the decoding matrix and answers with the $j$-th element of that row for the question $j$. 

Before we proceed further with finding the optimal (deterministic) strategies let us define the similarity between two strings as follows:
\begin{defn}[Similarity between two strings]
	We say that a string $x = x_1 x_2 \ldots x_n$  is ${\rm s}$-similar to a string $z = z_1 z_2 \ldots z_n$, denoted as \emph{$\mbox{sim}(x,z)={\rm s}$}, if $x$ and $z$ agree in exactly ${\rm s}$ positions. Formally, $\mbox{sim}(x,z)=\left|\left\{j:x_j=z_j\right\}\right|$.
\end{defn}
Given a decoding strategy $\mathcal{D}$ of Bob, the optimal encoding strategy of Alice is to send a letter $y_{\rm opt}\in X$ such that
\begin{equation}
	\mbox{sim}(x,\mathcal{D}_{y_{\rm opt}})\geq \mbox{sim}(x,\mathcal{D}_{y})~~\forall~~y\in\{0,\cdots, d\!-\!1\}, \label{eq:optenco}
\end{equation}
i.e., Alice send a $y_{\rm opt}$ such that the string "$\mathcal{D}\left(y_{\rm opt},1\right)~\mathcal{D}\left(y_{\rm opt},2\right)~ \ldots~ \mathcal{D}\left(y_{\rm opt},n\right)$" best approximates word $x=x_1~x_2~\ldots~x_n$ (of course, Alice can have more than one such optimal letter; in that case, she is free to choose any one of them). Therefore, for finding optimal classical protocols, it is sufficient to consider a smaller subset of deterministic protocols given by
\bea
\widetilde{\mathrm{S}}=\{\mathcal{S}=(\mathcal{E}^{\mathcal{D}}_{\rm opt},\mathcal{D}):~\mbox{where}~\mathcal{D}\in \mathbb{D}\},
\eea
where the notation $\mathcal{E}^{\mathcal{D}}_{\rm opt}$ is used to clarify the dependence of an optimal encoding on the decoding strategy of Bob. So from here onward we only consider the deterministic protocols of the following form
\begin{equation}
\mathcal{S}=(\mathcal{E}^{\mathcal{D}}_{\rm opt},\mathcal{D}),~~\mbox{where}~\mathcal{S}(x,j)= \mathcal{D}(\mathcal{E}^{\mathcal{D}}_{\rm opt}(x),j).
\end{equation} 
Since all the protocols in the subset $\widetilde{\mathrm{S}}$ depend only on the decoding matrices of Bob, they will be referred by their respective decoding matrices.  In a simplified notation, henceforth we represent decoding matrices of Bob as $f = [f_{yj}]$ where the row index $y\in\{0,...,d\!\!-\!\!1\}$, column index $j\in\{1,...,n\}$, and $f_{yj}\in X=\{0,...,d\!\!-\!\!1\}$, then decoding is represented by a $d\times n$ matrix of the form
\begin{equation}
	f = [f_{yj}]=\begin{bmatrix} 
		f_{01} & \dots  & f_{0n}\\
		\vdots & \ddots & \vdots\\
		f_{(d\!-\!1)1} & \dots  & f_{(d\!-\!1)n} 
	\end{bmatrix},
\end{equation}
and a protocol in $\widetilde{\mathrm{S}}$ is referred by a decoding matrix $f$.
The average success probability of any protocol $f$, can now be expressed as follows:
\bea
\mbox{Value}(f) &\equiv &\overline{\mathrm{P}}_{f}~~\mathrm{where} \nonumber \\
\overline{\mathrm{P}}_{f} &=& \frac{1}{d^n}\sum_{x\in X^n}\left[\max_y{\left(\frac{\mbox{sim}\left(x,f_{y\ast}\right)}{n}\right)}\right].
\eea
We are using notations: $f_{y\ast}$ to denote $y$-th row, $f_{\ast j}$ to denote $j$-th column, and $f_{yj}$ to denote $(y,j)$-th element of the matrix $f = [f_{yj}]$. Note that there is a total $d^{~\!nd}$ number of possible decoding matrices, our goal is to find decoding matrices that correspond to optimal protocols. Let us define the value of a word for a given strategy $f$ and the value of a strategy as follows

\begin{defn}[~Value of a word $x\in X^n$ for strategy $f$~ and value of the strategy]
	We define value of a word $x=x_1x_2...x_n$ for a strategy $f$ as 
	\begin{equation}
		v_{f}(x)=\frac{1}{d^n}\left[\max_y{\left(\frac{{\rm sim}\left(x,f_{y\ast}\right)}{n}\right)}\right]. \label{x-value}
	\end{equation}
	The value of the strategy $f$ is defined by adding the value of all the words, that is,
	\begin{equation}
		\mbox{Value}(f)=\sum_{x\in X^n} v_{f}(x). \label{eq:vof}
	\end{equation}
\end{defn}
In the following, we collect technical lemmas about decoding strategies and characterize the optimal classical strategies. The optimality is to be presented after addressing the relevant lemmas. Let us start with a property of an optimal decoding strategy. 
	
\begin{lem}\label{lemma001}
	Suppose two rows of a decoding matrix $f$ are identical. Replacing one of these identical rows with a new row that is not identical to any other row in $f$ will strictly increase the value of the strategy.
\end{lem}
	
	\begin{proof}[Proof]
		Suppose $y_1$ and $y_2$ rows of a decoding matrix $f$ are identical, i.e., $f_{y_1\ast} = f_{y_2\ast}$. Let us construct a transformation of the decoder, denoted by $\widetilde{f}$, by replacing the row $f_{y_2\ast}$ with another one, that is, a $n$-length string $\widetilde{f}_{y_2\ast}$ such that $\widetilde{f}_{y_2\ast}\neq f_{y\ast}$ for all $y$~\footnote{One can always find such a replacement because total number of $n$-length strings are $d^n$ whereas the number of rows in a decoding matrix is $d$.}. Then, Eq.~(\ref{x-value}) implies that $v_{\widetilde{f}}(x)\geq v_{f}(x)$ for all $x\in X^n$. Moreover, for the word $x=\widetilde{f}_{y_2\ast}$, we have strict inequality $v_{\widetilde{f}}(x)> v_{f}(x)$, from which we have $\mbox{Value}(\widetilde{f})>\mbox{Value}(f)$. 
	\end{proof}

A decoding matrix with no two rows equal may improve if it contains a column having a pair of identical elements; this essentially shows that making them no longer identical may allow for improving a decoder. 

 	\begin{lem}\label{lemma002} 
 Suppose a decoding matrix $f$ contains pairwise distinct rows and a column with two identical elements. Then, the value of the strategy may increase, but never decrease, by transforming the decoding matrix in the following way; replacing one of these identical elements in the column with a different element that does not match any other element in that column. 
	\end{lem}
	
	\begin{proof}[Proof]

Consider a decoding matrix $f$ with no two rows identical, and a column $j^{\prime}$ with two equal entries, say $f_{y_1j^{\prime}}=f_{y_2 j^{\prime}}=a$. Then there exists a letter $b$ which does not belong to the $j^{'}$ column of $f$.~\footnote{Since the column length is $d$, it has a repeated letter, and alphabet size is $d$, there is at least one alphabet $b$ distinct from all elements of the column.} Let us construct another decoding matrix $\widetilde{f}$ such that its elements fulfil, $\widetilde{f}_{y_2j^{\prime}}=b$, otherwise $\widetilde{f}_{yj}=f_{yj}$, see Fig.~\ref{fig1}. We aim to show that a revision of the decoding strategy may improve but never deteriorate, i.e., $\mbox{Value}(\widetilde{f})\geq \mbox{Value}(f)$.

The proof investigates the effect of revising a strategy, 
\bea
f \mapsto \widetilde{f} \label{eq:fftil}
\eea 
for all words in $X^n$, which may be divided into a few partitions. 

Let us begin by defining the set of words, denoted by $W_f$, for which $f_{y_2\ast}$ is an optimal row~\footnote{A row $y_{r}$ in a decoding matrix $\mathbf{f}$ is an optimal row for a word $x$ if~ $\mbox{sim}(x, \mathbf{f}_{y_r\ast})=\underset{y}{\max}~\mbox{sim}(x, \mathbf{f}_{y\ast})$, that is,
$$\mbox{sim}(x, \mathbf{f}_{y_r\ast}) \geq  \mbox{sim}(x,\mathbf{f}_{y\ast}) ~~\mbox{for~all~}~ y.$$}.
 Let us write by
\bea
W_f= \{x\in X^n~\vert~\forall y,~\mbox{sim}(x,f_{y_2\ast})\geq \mbox{sim}(x,f_{y\ast})\} \nonumber
\eea
and its complement as follows, 
\bea
 \overline{W}_f =  \{x \in X^n~\vert~  x\notin W_f\}. \nonumber 
\eea
Note that, by the above definition, the row $y_2$ in $f$ is not an optimal row for any word $x\in \overline{W}_f $.

	\begin{figure}[t!]
		\begin{center}
			\includegraphics[scale=0.25]{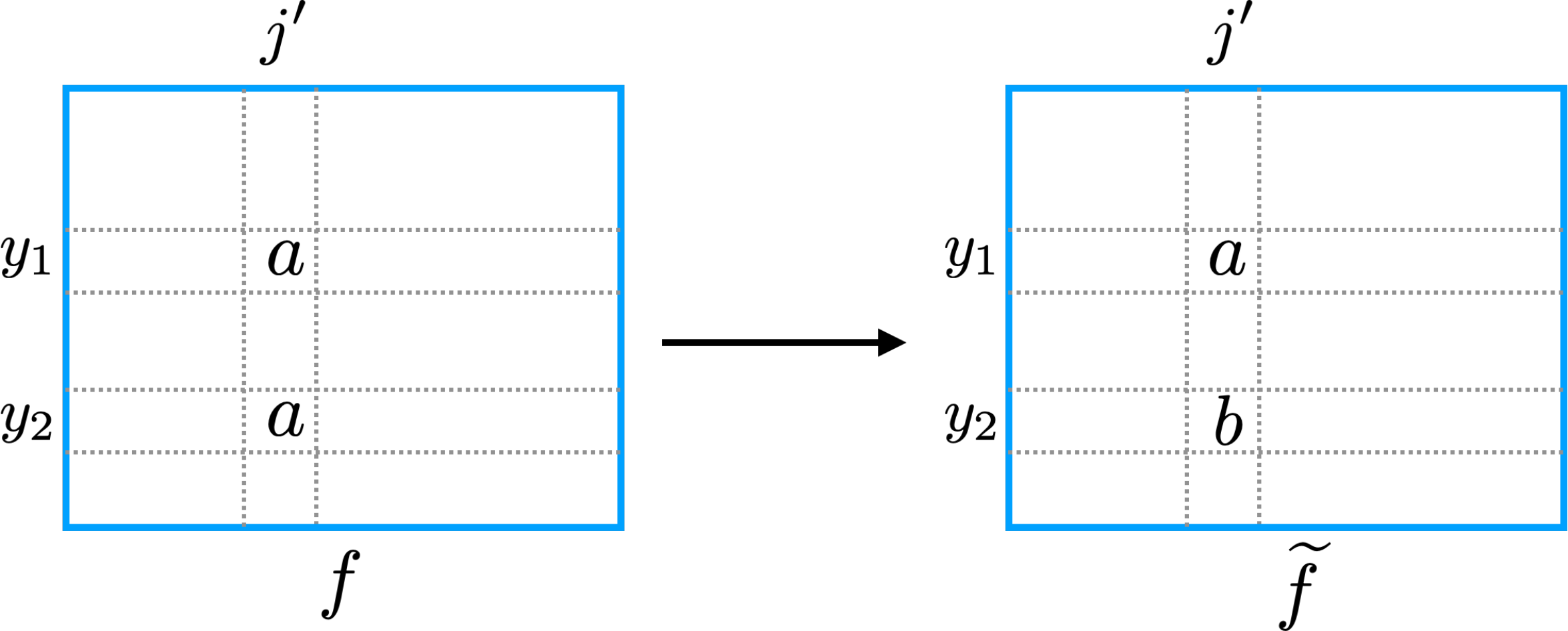}
			\caption{{\small A revision of a decoding strategy is attempted by replacing elements, $f \mapsto \widetilde{f}$.}}
			\label{fig1}
		\end{center}
	\end{figure}
 

The first instance is the set $\overline{W}_f$. For an $x\in \overline{W}_f$, there exists a row $y_{\rm opt}\neq y_2$ such that $\mbox{sim}(x,f_{y_{\rm opt}\ast}) \geq \mbox{sim}(x,f_{y\ast})$ for all $y$, by which we have
\bea
				v_f(x)=(nd^n)^{-1}~\mbox{sim}(x,f_{y_{\rm opt}\ast}). \label{eq:vfx}
\eea

For the same word $x\in \overline{W}_f$, on revising the decoding to $\widetilde{f}$ the value of the word $x$ may increase by at most one unit  [$1~\mbox{unit}~= (nd^n)^{-1}$] (see Fig.~\ref{fig1}), therefore we have
\bea
    \mbox{sim}(x,\widetilde{f}_{y_2\ast})\leq \mbox{sim}(x,f_{y_{\rm opt}\ast}).\label{eq:fopt}
\eea 
Collecting all these, we compute the value for a $x \in \overline{W}_f$ with a strategy $\widetilde{f}$ and find that $	v_{\widetilde{f}}(x) = v_{{f}}(x)$, see Eq.~(\ref{eq:vfx}): 
\bea
				v_{\widetilde{f}}(x) &=& (nd^n)^{-1}~\max_y ~\mbox{sim}(x,\widetilde{f}_{y\ast})\nonumber \\
				&=& (nd^n)^{-1}~ \max \left[\mbox{sim}(x,\widetilde{f}_{y_2\ast}),~\max_{y\neq y_2} ~\mbox{sim}(x,\widetilde{f}_{y\ast}) \right] \nonumber\\
				&=& (nd^n)^{-1}~ \max \left[\mbox{sim}(x,\widetilde{f}_{y_2\ast}),~\,\max_{y\neq y_2} ~\mbox{sim}(x,f_{y\ast}) \right] \nonumber\\
				&=&(nd^n)^{-1}~ \max \left[\mbox{sim}(x,\widetilde{f}_{y_2\ast}),~\mbox{sim}(x,f_{y_{\rm opt}\ast}) \right]\nonumber \\
				&=&(nd^n)^{-1}~\mbox{sim}(x,f_{y_{\rm opt}\ast}) = v_f(x) \nonumber
\eea
where we have used Eq. (\ref{eq:fopt}) to derive the third equality. The value remains identical by varying strategies in Eq. (\ref{eq:fftil}) for all $x\in \overline{W}_f$.

Then, let us consider the other set $W_f$, which we divide as follows. For a word $x=x_1~...~x_{j^{\prime}}~...x_n \in W_f$, we introduce
\bea
W^a_f, ~W^b_f, ~\mathrm{and}~ W^{\overline{ \{a,b \} }}_f, \nonumber
\eea
such that it belongs to $W_{f}^{ x_{j^{\prime}}}$ for $x_{j^{\prime}}=a$, $x_{j^{\prime}}=b$, and $x_{j^{\prime}}\notin \{a,b\}$, respectively. Note that, by definition, the row $y_2$ in $f$ is an optimal row for all the words $x\in W_f $.  

As the second instance, let us consider a word $x\in W^{\overline{ \{a,b \} }}_f$. 
Since $x_{j^{\prime}}\notin \{a,b\}$, it is easy to see that $v_{\widetilde{f}}(x)=v_{f}(x)$ for all these words (see Fig.~\ref{fig1}).

The third instance is the set $W^{b}_f$ in which the value increases, that is, $v_{\widetilde{f}}(x)=v_{f}(x)+1$. To distinguish this case from the others, let us also call an element in the set a $b$-word. 

As the fourth instance, let us consider the words in $x\in W^{a}_f$, which can also be classified into two parts $A_f$ and its complement $\overline{A}_f$. (i) $A_f$ consists of words for which no other row of $f$, except the $y_2$-th row, is optimal, i.e.,
\bea
A_f=\{x\in W^{a}_f~\vert~ \mbox{sim}(x,f_{y_2\ast})>\mbox{sim}(x,f_{y\ast})~~\forall~y\neq y_2\}. \nonumber 
\eea
(ii) $\overline{A}_f$ consists of words for which two or more rows (including the $y_2$-th row) are optimal,
\bea
\overline{A}_f= \{x \in W^{a}_f~\vert~  x\notin A_f\}.
\eea
For all $x\in \overline{A}_f$, it holds  that $v_{\widetilde{f}}(x)=v_{f}(x)$. Next, we consider the set $A_f$, where any word fulfills $v_{\widetilde{f}}(x)=v_{f}(x)-1$ and is called an $a$-word.

It is clear that $A_f$ is nonempty since no two rows in $f$ are identical. For example, $f_{y_2\ast} \in A_f$. We show that there is an injective map from $A_f$ to $W^{b}_f$, i.e., an $a$-word to a $b$-word. Let $x^a$ denote an element, $x^a=x_1~...~a~...~x_n \in A_f$, and then it holds, 
   \bea
   \mbox{sim}(x^a,f_{y_2\ast})>\mbox{sim}(x^a,f_{y\ast})~~\forall~y\neq y_2.\nonumber
   \eea
For a word $x^a$ corresponds a word $w=x_1~...~b~...~x_n$, and $\mbox{sim}(w,f_{y_2\ast})\geq \mbox{sim}(w,f_{y\ast})~~\forall~y\neq y_2$ implying $w\in W^{b}_f$. Thus $x^a\mapsto w$ is an injective map from the set $A_f$ to the set $W^{b}_f$, therefore $|W^{b}_f|\geq |A_f|$. 

	\begin{figure}[t!]
		\begin{center}
			\includegraphics[scale=0.25]{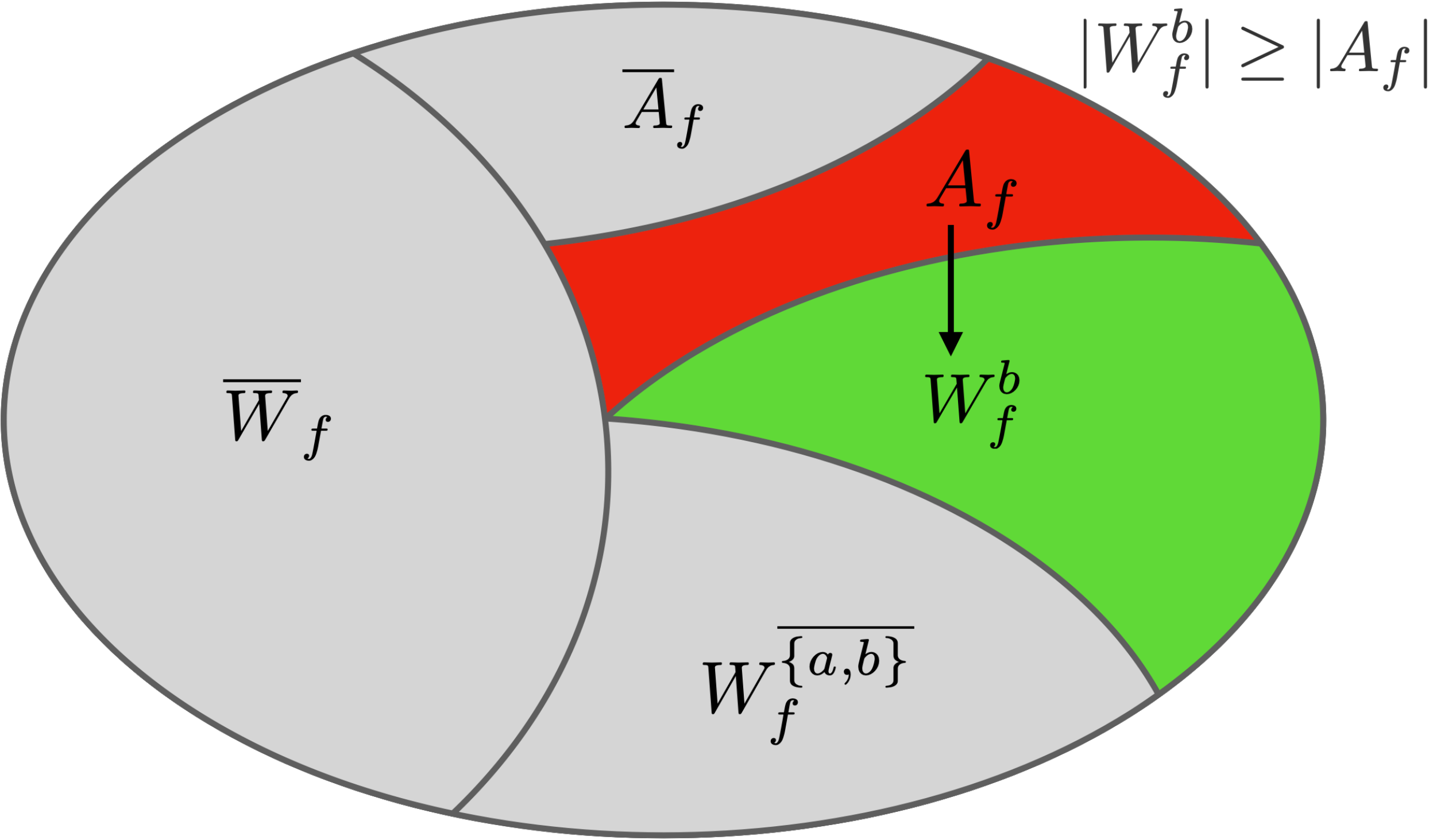}
			\caption{{\small Partitions of the set of all words $X^n$. The effect of revising the decoding $f \rightarrow \widetilde{f}$ are as follows: (i) does not affect the parts $\overline{W}_f$, $W^{\overline{\{a,b\}}}_f$, and $\overline{A}_f$ (regions shaded grey), (ii) the value of each word in $W^b_f$ is increased by one unit (region shaded green), (iii) the value of each word in $A_f$ is decreased by one unit (region shaded red), (iv) there is an injective map $A_f \mapsto W^b_f$ implying $\vert W^b_f\vert \geq \vert A_f\vert$.}}
			\label{fig2}
		\end{center}
	\end{figure}
 

  To conclude, we have investigated all elements in $X^n$, comprised of five parts, $\overline{W}_f$, $W^{\overline{\{a,b\}}}_f$, $W^b_f$, $A_f$, and  $\overline{A}_f$ (see Fig.~(\ref{fig2})). For all the words in $\overline{W}_f$, $W^{\overline{\{a,b\}}}_f$, $\overline{A}_f$, the value remains the same while a strategy varies in Eq. (\ref{eq:fftil}), whereas it decreases by one unit for every word in $A_f$ and increases by one unit for for each word in $W^{b}_f$ [1 unit $=(nd^n)^{-1}$]. Moreover, we have shown that, there is an injective mapping from $A_f$ to $W^{b}_f$. Thus, we conclude that $\mbox{Value}(\widetilde{f})\geq \mbox{Value}(f)$.
	\end{proof}
	
So far, two desired properties of decoding matrices have been established. These properties provide an ordered structure for comparing two decoding strategies, ultimately leading to the identification of an optimal decoding matrix. Starting from an arbitrary initial decoding strategy, if there are identical rows, one can apply Lemma~\ref{lemma001} repeatedly to achieve a decoding strategy in which no two rows are identical. Each such transformation results in a strict increase in the value of the decoding strategy. The resulting strategy may still have columns with repeated entries. Next, by applying Lemma~\ref{lemma002} successively, without decreasing the value of the decoding strategy at any step, one can gradually revise the strategy so that every column contains distinct elements. Let $F^{\star}$ denote the set of decoding strategies that satisfy this property--namely, that all elements in each matrix column are distinct. We will now prove that this set characterizes the optimal strategies.

For instance, the so-called \emph{majority-encoding-identity-decoding} (MEID) strategy in Ref. \cite{Tavakoli} is also in the set $F^{\star}$. In this case, a decoding matrix is given by $f^{\rm maj}_{yj}=y$, i.e.,
\begin{equation}
 f^{\rm maj}=\begin{bmatrix} 
			0 & \dots  & 0 \\
			\vdots & \ddots & \vdots\\
			d\!-\!1 & \dots  & d\!-\!1 
		\end{bmatrix}.
\end{equation}
Note that, in MEID, the majority rule is applied as an optimal encoding by Alice; she sends a letter that is most frequent in her input word $x=x_1~...~x_n$. In cases where more than one letter is equally frequent, Alice can choose one of them. In the following lemma we show that it cannot be improved anymore.

	\begin{lem}\label{lemma003}
A strategy in the set $F^{\star}$ is optimal. 			
	\end{lem}
	
	\begin{proof}[Proof]
		Consider a strategy $f^{\star}\in F^{\star}$. For each column $j$, all letters $f^{\star}_{yj}$, where $y \in X$, are different, so, for $j\in \{1,~...~,n\}$ columns of $f^{\star}$ are given by $f^{\star}_{\ast j}=[\sigma_j(0),\sigma_j(1),~...~,\sigma_j(d\!\!-\!\!1)]^T$ where $\sigma_j$ are permutation functions defined over the alphabet $X=\{0,1,~...~,d\!\!-\!\!1\}$. Then a decoding matrix $f^{\star}\in F^{\star}$ can be expressed as
		\begin{equation}
			f^{\star} = [f^{\star}_{yj}]=\begin{bmatrix} 
				\sigma_1(0) & \dots  & \sigma_n(0)\\
				\vdots & \ddots & \vdots\\
				\sigma_1(d\!\!-\!\!1) & \dots  & \sigma_n(d\!\!-\!\!1)\\
			\end{bmatrix}.
		\end{equation}
		
		If $f^{\star}=f^{\rm maj}$ then $\mbox{Value}(f^{\star})=\mbox{Value}(f^{\rm maj})$. Let us consider $f^{\star}\neq f^{\rm maj}$, then, we note that the value $v_{f^{\rm maj}}(x)$ of a word $x = x_1 \ldots x_n$ with the strategy $f^{\rm maj}$, is same as the value $v_{f^{\star}}(x^{\star})$ where $x^{\star} = \sigma_1(x_1) \ldots \sigma_n(x_n)$. Since permutation maps are invertible, the defined map $x\mapsto x^{\star}$ is one-to-one and onto. Therefore, $\mbox{Value}(f^{\star})=\sum_{x^{\star}}v_{f^{\star}}(x^{\star})=\sum_{x}v_{f^{\rm maj}}(x)=\mbox{Value}(f^{\rm maj})$. 
		
		The optimality of all the $f^{\star}$ strategies follows from the Lemma~\ref{lemma002}. Any modification to a $f^{\star}$ strategy such that it is mapped out of the set $F^{\star}$ results in a decoding matrix in which at least one letter will appear twice in at least one of the columns. Then Lemma~\ref{lemma001} and Lemma~\ref{lemma002} implies that any such revision cannot increase the value of the strategy. Note that, starting from an arbitrary initial strategy, we can reach global maxima through a sequence of non-decreasing transformations.
	\end{proof}

 Having established the lemmas above, we are now prepared to characterize optimal classical strategies.

\begin{thm}\label{theorem001c}
 For all $(n,d)\!\mapsto\!1$ classical random access codes (RACs), a deterministic strategy is optimal if all elements in each column of the decoding matrix are distinct and , given such a decoding, an optimal encoding in Eq~(\ref{eq:optenco}) is implemented. An example of such an optimal strategy is the MEID strategy.
\end{thm}

The above-stated theorem is extended to provide both necessary and sufficient conditions for optimality. Details of the analysis are presented in \ref{appendix-A}.
	 
\subsection{Optimal classical values}

Having proved that the MEID strategy is optimal, in this section we implement the algorithm that applies the strategy to compute an optimal success probability. The algorithm also reproduces results in Ref. \cite{Tavakoli}. It is explicitly constructed as follows. 

\begin{algorithm}[h]
    \caption{Computing Optimal Classical Value}
    \textbf{Input:} N $\leftarrow$ upper bound on word length\\
    \textbf{Input:} D $\leftarrow$ upper bound on alphabet size
    
   DEFINE M, initialized to 0, as a two-dimensional array of size $N \times D$\;
    
    \For{n $\leftarrow$ 2 \KwTo N}{
        \For{d $\leftarrow$ 2 \KwTo D}{
            VAL $\leftarrow$ 0\;
            
            \For{k $\leftarrow$ 0 \KwTo $d^n - 1$}{
                LetterCounts $\leftarrow$ CountLetters(k, d, n) \; \tcp{CountLetters(k,d,n) counts letters in a n digit number k represented in base d}
                VAL $\leftarrow$ VAL + $\frac{1}{n \cdot d^n} \cdot \text{Max(LetterCounts)}$\;
            }
            M[n][d] $\leftarrow$ VAL\;
        }
    }
    
    \textbf{Output:} M\;
\end{algorithm}

Applying the algorithm, we have obtained optimal probabilities for $d\in\{2,3,4,5,6\}$ and $n\in\{2,3,4,5,6\}$, which are shown in Table~\ref{table1}. It is worth mentioning that the results for $n=4,5,6$ and $d>2$ for which optimal probabilities are not known are computed.  
 \begin{center}
		\begin{table}[h!]
			\centering
			\begin{tabular}[b]{c||ccccc } 
				${\rm n}$~$\backslash$~${\rm d}$ & 2 & 3 & 4 & 5 & 6 \\ [3pt]
				\hline \hline \\
				2~~~~~ & ~$\frac{3}{4}$~ & ~$\frac{2}{3}$~ & ~$\frac{5}{8}$~ & ~$\frac{3}{5}$~ & ~$\frac{7}{12}$~\\ [5pt]
				3~~~~~ & ~$\frac{3}{4}$~ & ~$\frac{17}{27}$~ & ~$\frac{9}{16}$~ & ~$\frac{13}{25}$~ & ~$\frac{53}{108}$~\\ [5pt]
				4~~~~~ & ~$\frac{11}{16}$~ & ~$\frac{16}{27}$~ & ~$\frac{17}{32}$~ & ~$\frac{61}{125}$~ & ~$\frac{197}{432}$~ \\ [5pt]
				5~~~~~ & ~$\frac{11}{16}$~ & ~$\frac{5}{9}$~ &~ $\frac{127}{256}$~ & ~$\frac{1429}{3125}$~ &~ $\frac{185}{432}$~\\ [5pt]
				6~~~~~ & ~$\frac{21}{32}$~ & ~$\frac{131}{243}$~ &~ $\frac{481}{1024}$~ &~ $\frac{1341}{3125}$~ &~ $\frac{3121}{7776}$~\\ [5pt]
				\hline 
			\end{tabular}\\[1pt]
			\caption{{\small The value of maximum success probability in classical RACs when $n,d\in\{2, 3, 4, 5, 6\}$.}\label{table1}}
		\end{table}
	\end{center}

In addition, the algorithm reproduces optimal probabilities known from analytic expressions. For instance, in Ref.~\cite{Tavakoli}, the probabilities are known respectively for $n=2$ and $n=3$ with $d\geq 2$, 
	\begin{eqnarray}
		 \overline{\mathrm{P}}_{cl}&=&\frac{1}{2}\left\{ 1+\frac{1}{d}\right\}, \\  
		\overline{\mathrm{P}}_{cl}&=&\frac{1}{3}\left\{1+\frac{3}{d}-\frac{1}{d^2}\right\}.
	\end{eqnarray}
In Ref.~\cite{Ambainis2008}, the optimal probability has been obtained for $n\geq2$ and $d=2$, 
	\begin{align}
		\overline{\mathrm{P}}_{cl}=\frac{1}{2}+\frac{1}{2^n}~\binom{n-1}{\lfloor\frac{n-1}{2}\rfloor},  
	\end{align}
	where $\binom{{\rm n}}{{\rm k}}$ denotes the ${\rm n}$-choose-${\rm k}$ function, and $\lfloor{\rm x}\rfloor$ is the floor function. The algorithm is consistent to the closed-form of the optimal probabilities above. \\

\section{Quantum advantage}\label{sec:4}
In this section, we investigate quantum strategies and show quantum advantages in RACs. We construct QRAC protocols with minimal resources, two incompatible measurements.


 \subsection{QRAC with a single measurement}

 A QRAC that exploits quantum resources encodes a word $x=x_1,~...,~x_n\in X^{n}$ in a quantum state on a $d$-dimensional Hilbert space. Bob then applies a measurement, denoted by a POVM $M = \{\Pi_j\}^{d-1}_{j=0}$, for decoding. 
	\begin{thm}  
	A QRAC protocol using a single measurement setting in decoding does not give an advantage over classical strategies. 
	\end{thm}
	\begin{proof} 
Let Alice's encoding denote $x\mapsto \rho_x$, a quantum state in a $d$-dimensional Hilbert space. Suppose Bob applies only a single measurement for decoding as follows,  
\bea
\mathcal{M} = \{\Pi_{0}, \Pi_{1}, \cdots, \Pi_{d-1} \} \label{eq:mm}
\eea
where $\Pi_{\kappa}$ are POVM elements and $\sum^{d\!-\!1}_{\kappa=0}\Pi_{\kappa}=\mathbbm{1}$. 

We write a quantum protocol with only one measurement by $Q^0$. Referring to Eq. (\ref{eq:gue}), suppose that Bob chooses $j\in \{1,\cdots,n \}$ with a probability $1/n$, meaning he wants to learn a letter $x_j$ in a word $x$. A complete measurement in Eq. (\ref{eq:mm}) is applied and the POVM element giving an outcome $x_j$ is denoted by $\Pi_{x_j}$; hence, the success probability is given by ${\rm tr}[\rho_x \Pi_{x_j}]$. Then, the average success probability for a protocol $Q^0$ is computed as follows,
	\begin{equation}
		{\rm Value}(Q^0)=\frac{1}{nd^n}\sum_{x\in X^n}~\mbox{tr}~[~\rho_x(\Pi_{x_1}+\Pi_{x_2}+...+\Pi_{x_n})].\nonumber
	\end{equation}
The average success probability is maximal when $\rho_x$ is an eigenstate from the maximal eigenvalue of the operator $(\Pi_{x_1}+\Pi_{x_2}+...+\Pi_{x_n})$ \cite{Carmeli_2020,FarkasRACs1}. 

It is thus clear that a state $\rho_x$ which gives a maximum success probability is of rank-one; this naturally corresponds to the computation of an operator norm, denoted by $\| \cdot \|_{op}$, $\|A \|_{op} = \max_{\||v\rangle\|=1} \| A |v\rangle \|$. That is,
\bea
\max_{\rho_x} {\rm Value}(Q^0) = \frac{1}{nd^n}\sum_{x\in X^n} \| \Pi_{x_1}+ \cdots +\Pi_{x_n} \|.\nonumber
\eea
The right-hand-side for a word $x$ is bounded above, 
\bea
\| \Pi_{x_1}+ \cdots +\Pi_{x_n} \| & = & \|c_0 \Pi_{0}+ \cdots + c_{d-1} \Pi_{x_{d-1}} \| \nonumber \\
& \leq  & m_x \| \Pi_{0}+ \cdots +  \Pi_{d-1} \|, \nonumber 
\eea
where $c_j$ is the number of a letter $j\in\{0,\cdots,d-1 \}$ in a word $x=x_1\cdots x_n$ and $m_x = \max_{j} c_j$. Hence, we have
\bea
\max_{\rho_x} {\rm Value}(Q^0) \leq \frac{1}{nd^n}\sum_{x\in X^n} {m_x} \label{eq:cmax}
\eea
which can be achieved by classical resources, such as the MEID strategy. Thus, we have shown that a QRAC protocol with a single measurement does not go beyond classical optimal values. 

\end{proof}
We note that alternative proofs of our Theorem-2 may follow from the Frenkel-Weiner theorem~\cite{Frenkel-Weiner} or results proving the necessity of incompatible measurements for showing advantage in a class of general communication tasks~\cite{Deba+PRA2023}  or in particular for the random access codes~\cite{Carmeli_2020}.
	
\subsection{QRAC with two measurements}

We here present a QRAC protocol that applies two measurements for decoding and show that it achieves a higher value than classical optimal strategies. 

\subsubsection{The protocol}
The QRAC with two measurement settings applies two mutually unbiased bases (MUBs), the computational and the Fourier ones, 
\bea
 \{ | 0\rangle, \cdots, |1 \rangle \}~~\mathrm{and}~~   \{ |  e_0\rangle,\cdots, | e_{d-1} \rangle \}\nonumber 
\eea
where 
\bea
\vert e_l\rangle=\frac{1}{\sqrt{d}}~\sum_{k=0}^{d\!-\!1}~\omega^{kl}~\vert k\rangle~~ {\rm and}~~\omega=\exp\left(2 \pi i/d\right). \nonumber
\eea
In the following, we present a QRAC protocol with two MUBs for decoding and show the optimization of the encoding of Alice. 

Alice begins encoding of a word $x=x_1,~...,~x_n$ by finding parameters as follows. 
\begin{itemize}
    \item $L_x$: a set of the most frequent letters in $x$, and 
    \item $m_x$: the number of occurrences of an element of $L_x$ for $x$. 
\end{itemize}
For instance, for $x=001233$ for $n=6$, we have $L_x=\{ 0,3\}$ and $m_x=2$ since letters in $L_x$ appear twice, respectively. It is clear that $1\leq | L_x| \leq |X|$ where $|A|$ denotes the cardinality of a set $A$. 


If \( x_n \in L_x \) and \( |L_x| > 1 \), Alice introduces a new set \( L_x' \) by excluding \( x_n \) from \( L_x \), so \( x_n \notin L_x' \). If \( L_x \) is a singleton, meaning \( L_x = \{ x_n \} \), then \( L_x' \) remains equal to \( L_x \). On the other hand, if \( x_n \notin L_x \), Alice keeps \( L_x' \) equal to \( L_x \). Therefore, we have:

\[
L_x' = \begin{cases}
\{ x_i \in L_x \mid x_i \neq x_n \} & \text{if } x_n \in L_x \text{ and } |L_x| > 1, \\
L_x & \text{otherwise}.
\end{cases}
\]

With the revised set above, Alice can prepare a quantum state for a word $x$. She chooses an element, denoted by $l^{\star}$, from the set $L_{x}^{'}$ and creates a superposition of two states, a computational basis $|l^{\star}\rangle$ and a Fourier basis $|e_{x_n}\rangle$, to encode a word $x$ as follows, 
\bea
	 |\psi_x\rangle&\!=\!&{\rm N}_{x}~ \left( \alpha_x~|l^{\star}\rangle +~ \beta_x~\exp(i\phi_x)~ |e_{x_n}\rangle\right), ~\label{eq:stt}
\eea
where parameters $(\alpha_x, \beta_x,\phi_x)$ are to be optimized for obtaining quantum advantages, and ${\rm N}_x$ is a normalization factor
 \bea
 {\rm N}_{x}&=&\left[\alpha_x^2 + \beta_x^2 + \frac{2\alpha_x \beta_x \lambda_x}{\sqrt{d}}\right]^{-\frac{1}{2}},~~\mbox{where}  \\ [4pt]
 \lambda_x &=& \cos\left(\!\phi_x + \frac{2\pi l^{\star}x_n}{d} \right). \nonumber \label{eq:Nx}  
 \eea 
One can observe that to create a superposition, Alice selects a state $|l^{\star}\rangle$ from the computational basis and the other one $|e_{x_n}\rangle$ in a Fourier basis is automatically chosen from the last letter of a given word $x$.

Bob applies a measurement for decoding. The decoding strategy is set by two measurement settings. He applies projective measurements in the computational basis to guess the first $n\!-\!1$ letters $\{x_j: ~j\in \{1,~...~,n\!-\!1\}\}$ and in the Fourier basis for guessing the last letter $x_n$. Bob then reports a measurement outcome as his guess for the letter $x_j$.

 \subsubsection{Optimization of parameters: quantum advantages }

We here analyze the protocol and optimize the parameters ($\alpha_x, \beta_x,\phi_x$) for a word $x$. We refer to Eq. (\ref{eq:gue}) to compute the value for the QRAC protocol with two measurements. Let us denote the strategy with our quantum protocol by $\mathcal{S}_{Q}(x,j)$ and the guessing probabilities on applying the protocol by 
\bea
\mathrm{P}_Q(x,j)\equiv \mathrm{P}[\mathcal{S}_{Q}(x,j) \!=\!x_j]. \nonumber \label{eq:Pq}
    \eea
 Then, for a word $x$, the probability of a correct guess for letters at different positions is computed as follows: 
\begin{enumerate}
    \item  $\mbox{for}~ j \! \leq\!n\!-\!1$ and $x_j \!\neq\! l^{\star}$
    \bea
\mathrm{P}_{Q}^{(1)} (x,j) = \!{\rm N}_{x}^2 \left(\frac{\beta_x^2}{d}\right) \nonumber
    \eea
    \item  $\mbox{for}~ j \! \leq\!n\!-\!1$ and $x_j \! = \! l^{\star}$
    \bea
\mathrm{P}_{Q}^{(2)} (x,j) = 		\!{\rm N}_{x}^2 \left( \alpha_x^2\!+\!\frac{\beta_x^2}{d}\!+\!\frac{2\alpha_x\beta_x}{\sqrt{d}}\lambda_x\right),~~ \mathrm{and}  \nonumber 
    \eea
        \item  $\mbox{for}~ j= n$
    \bea
\mathrm{P}_{Q}^{(3)}(x,j) = 	\!{\rm N}_{x}^2 \left ( \beta_x^2\!+\!\frac{\alpha_x^2}{d}\!+\!\frac{2\alpha_x\beta_x}{\sqrt{d}}\lambda_x\right), \nonumber 
    \eea
\end{enumerate}
where $\mathrm{N}_x$ is in Eq. (\ref{eq:Nx}). The quantum value for a word $x$ also depends on whether $x_n$ is equal to $l^{\star}$, or not. For clarification, let us define a parameter, for a selection of letter $l^{\star} \in L_{x}^{'}$ in the encoding, 
	\begin{eqnarray}
		\epsilon_x
		= \begin{cases}
			1~& \mbox{if}~~~~x_n= l^{\star}, \\[5pt]
			0~&  \mbox{if}~~~~x_n\neq  l^{\star}. 
		\end{cases} 
	\end{eqnarray}
Then, the quantum value of a word $x$ for the protocol $Q$ is defined as average probability of correct answer by Bob over the set of all possible questions $j\in \{1,\cdots,n\}$, and it can be computed from the expression 
\bea
 	v_{Q}(x) & = &\frac{1}{nd^n} \times 
  \left[ (n\!-\!1\!)~\mathrm{P}_{Q}^{(1)}+ (m_x -\epsilon_x)~(\mathrm{P}_{Q}^{(2)}-\mathrm{P}_{Q}^{(1)})  + \mathrm{P}_{Q}^{(3)} \right].
  \label{eq:Qval}
\eea 
To find if a quantum advantage occurs, we are interested in the difference between $v_{Q}$ above and the corresponding value that can be achieved classically, see Eq. (\ref{eq:cmax}),
\bea
 	v_{\Delta Q}(x)&:=&v_{Q}(x)- \frac{1}{nd^n}m_x. \label{eq:diff} 
\eea
To compute the difference, we consider two classes, $x_n \neq l^{\star}$ or $x_n = l^{\star}$. In the former, we have $\epsilon_x=0$ and call $x^0$-words. For the latter, we have $\epsilon_x=1$ and write by $x^1$-words.

Quantum advantages are obtained if the difference $v_{\Delta Q}$ in Eq. (\ref{eq:diff}) is positive. For each word, we compute:  
\bea
	v_{\Delta Q}(x^0) \!&=&\! \frac{1}{nd^n}~\frac{\gamma_{x^0} ^2+2~\gamma_{x^0} ~\sqrt{d} ~\lambda_{x^0} -\kappa_{x^0} }{\left(\gamma_{x^0} ^2+1\right) d+2 ~\sqrt{d}~\gamma_{x^0}~\lambda_{x^0} },  \\[5pt]
v_{\Delta Q}(x^1)) \!&=&\!-\frac{1}{nd^n}~\frac{\gamma_{x^1} ^2 ~(d-1)+\kappa_{x^1} }{\left(\gamma_{x^1} ^2+1\right) d+2~ \sqrt{d} ~\gamma_{x^1}~  \lambda_{x^1} },
\eea
where $\kappa_{x^{\iota}}=d m_{x^{\iota}}+1-d-n$ and $\gamma_{x^{\iota}}=\alpha_{x^{\iota}}/\beta_{x^{\iota}}$ for $\iota\in \{0,1\}$. 
We attempt to find optimal $(\alpha_x, \beta_x, \phi_x)$ to maximize the difference $v_{\Delta Q}(x)$. With the help of Mathematica, we exploit an ansatz $\lambda_x=1$, which means, 
\bea
\phi_x=-\frac{2\pi l^{\star}x_n}{d}.\nonumber
\eea
With the ansatz, we choose the remaining parameter $\gamma_x\equiv (\alpha_x, \beta_x)$ for two cases. 
For a $x^0$-word, we choose the variable $\gamma_{x^0}$ as 
\begin{eqnarray}
\!\!\!\!\!\!\!\!\!\!\!\!&~&\gamma_{x^0} =\frac{\sqrt{\frac{1}{2} d (2 d+\kappa_{x^0}  (\kappa_{x^0} +\xi_{x^0} +4)+\xi_{x^0} -1)-\kappa_{x^0} }}{d-1}\! , \\[8pt]
\!\!\!\!\!\!\!\!\!\!\!\!&~&\mbox{where}~\xi_{x^0}=\sqrt{-\frac{4 \kappa_{x^0} }{d}+4 d+\kappa_{x^0}  (\kappa_{x^0} +6)-3},
\vspace{4cm}
\label{qv2}
\end{eqnarray}
and for a $x^1$-word, we have 
\begin{eqnarray}
	\!\!\!\!\!\!\!\!&~&\gamma_{x^1} \!\!=\!\!\begin{cases}
			\!\!\sqrt{\frac{d \left(d^2-d (\xi_{x^1} + 2)+ \kappa_{x^1} ^2+ \kappa_{x^1}  \xi_{x^1} +\xi_{x^1} +1\right)}{2 (d-1)^2}\!-\!\kappa_{x^1} }& \!\!\mbox{if}~\kappa_{x^1}\! >\! 0 \\[8pt]
			\!\!0&  \!\!\mbox{else}, 
	\end{cases} , \\[8pt]
	\!\!\!\!\!\!\!\!&~&\mbox{where}~\xi_{x^1}\!=\!\sqrt{d^2\!-\!2 d (\kappa_{x^1} \!+\!1)\!-\!\frac{4 \kappa_{x^1} }{d}\!+\!\kappa_{x^1}  (\kappa_{x^1} \!+\!6)\!+\!1}. 
	\label{qv2}
 \vspace{2cm}
\end{eqnarray}
Together with the normalization condition in Eqs. (\ref{eq:stt}) and (\ref{eq:Nx}), it is straightforward to find a pair $(\alpha_x, \beta_x)$.

Having identified the parameters $(\alpha_x,\beta_x,\phi_x)$ above, let us write the figure of merit, 
\bea
 	\mbox{Adv}(\mathcal{S}_Q)&=&\sum_{x\in X^n} v_{\Delta Q}(x) \label{eq:qa}
\eea 
where $ S_Q$ denotes the quantum strategy with two measurements. One can compute the above by 
\bea
\sum_{x^0\in X^n} v_{\Delta Q}(x^0)+\sum_{x^1\in X^n} v_{\Delta Q}(x^1).
\eea
 Quantum advantages can be assured if $\mbox{Adv}(\mathcal{S}_Q)>0$. 
 

\subsection{Examples}

We here consider the advantage of a quantum strategy by computing ${\rm Adv}(S_Q)$ in Eq. (\ref{eq:qa}) for cases $n=2,3,4$ and finite $d$. 

\subsubsection{Case $\mathbf{n=2}$ and $\mathbf{d\geq 2}$~.--}

For $n=2$, we have $m_x \in \{1,2\}$. The number of words having $m_x=1$ is $d^2-d$, which are of type $x^0$. In this case, we have $\kappa_{x^0}=-1$ and 
\bea
\xi_{x^0}=\sqrt{4/d +4d-8}. \nonumber
\eea
The number of words with $m_x=2$ is $d$, which are of type $x^1$. In this case, we have $\kappa_{x^1}=d-1$ and 
\bea
\xi_{x^1}=\sqrt{(12d^2+4(d-1)^2)/d}.\nonumber
\eea
After all, we have $\gamma_{x^0}=\gamma_{x^1}=1$, and thus,
\begin{eqnarray}
		v_{\Delta Q}(x^0) &=&\frac{1}{2d^2}\left( \frac{1}{\sqrt{d}}\right),~~\mathrm{and}\\
		v_{\Delta Q}(x^1)&=&\frac{1}{2d^2}\left(-1+\frac{1}{\sqrt{d}}\right).
\end{eqnarray}
For both types of words, the quantum strategy with two measurement settings has a quantum value
\bea
v_Q(x)=\frac{1}{2d^2} \left(1+ \frac{1}{\sqrt{d}}\right). \nonumber 
\eea
The average success probability is computed from Eq.~(\ref{eq:gue}),
\begin{equation}
\overline{\mathrm{P}}_{Q}=\frac{1}{2}\left( 1+\frac{1}{\sqrt{d}}\right).\label{eq:qopt2} 
\end{equation}
which is greater than the classical optimal one
\begin{equation}
	\overline{\mathrm{P}}_{cl}=\frac{1}{2}\left\{1+\frac{1}{d}\right\}.\nonumber
\end{equation}
The quantum advantages in a $(2,d)\!\mapsto\! 1$ RAC may be quantified as,
\begin{equation}
 	\mbox{Adv}(\mathcal{S}_Q)= \frac{1}{2}\left\{\frac{1}{\sqrt{d}}-\frac{1}{d}\right\}.\label{eq:qa2}
 \end{equation}
which is positive for all $d\geq 2$. The Eq. (\ref{eq:qa2}) has also been reported in \cite{Tavakoli} by introducing a quantum protocol, and based on a conjecture that the MEID protocol leads to the optimal classical value~\cite{Tavakoli} which we showed in the Sec.~\ref{sec:3} is indeed an optimal one. The reported quantum value in Eq. (\ref{eq:qopt2}) is the optimal value was proven later in the Ref.~\cite{FarkasRACs1}.

\subsubsection{Case $\mathbf{n=3}$ and $\mathbf{d}\geq 2$~.--} 

\begin{figure}[t!]
     \centering
     \begin{subfigure}[t]{0.42\textwidth}
         \centering
         \includegraphics[scale=0.5]{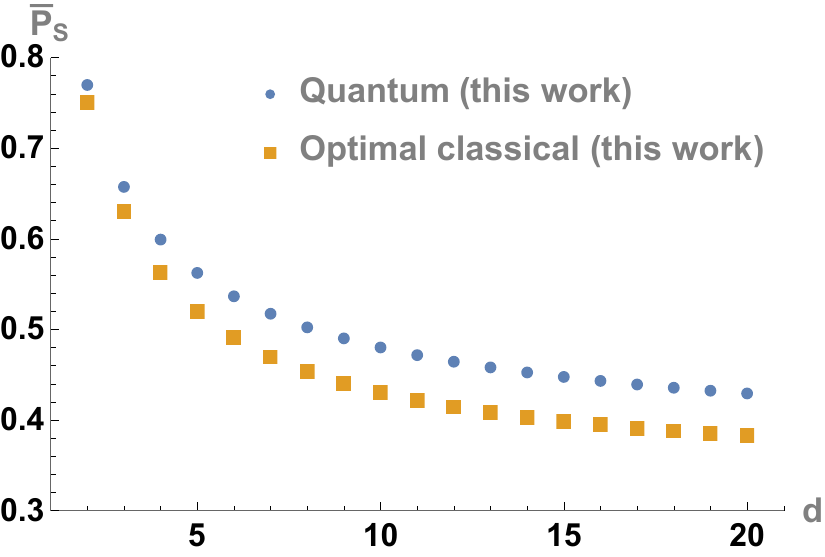}
         \caption{{\small Quantum and optimal classical probabilities in Eqs. (\ref{qf3d}) and (\ref{cf3d}) are shown for $d\leq 20$, in which quantum advantages are achieved.}}
          \vspace{0.5cm}
         \label{figa_3dRAC}
     \end{subfigure}
     \hfill
     \begin{subfigure}[t]{0.42\textwidth}
         \centering
         \includegraphics[scale=0.5]{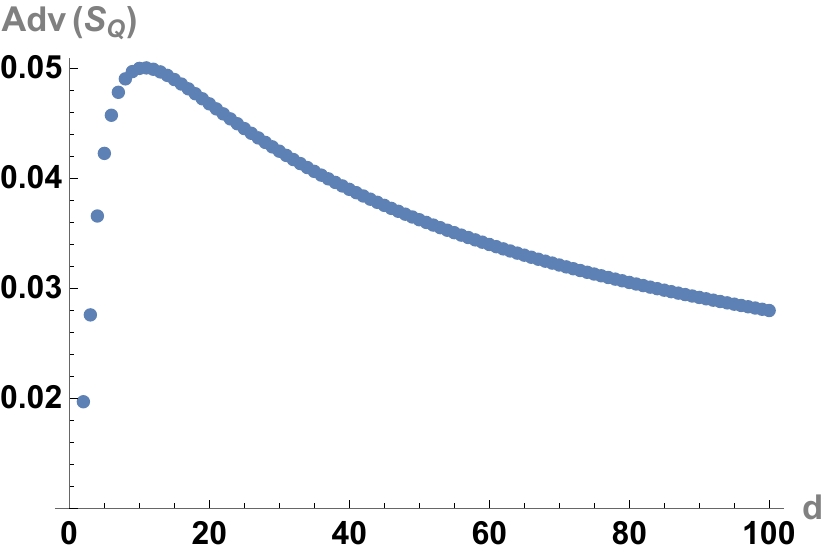}
         \caption{{\small The difference $\mbox{Adv}(\mathcal{S}_Q)$ is computed up to $d\leq 100$.}}
         \label{figb_3dRAC}
     \end{subfigure}
        \caption{ {\small Quantum advantages in $(3,d)\!\mapsto\! 1$ RACs for $d\leq 100$ are shown. The proposed quantum protocol applies two MUBs and shows a success probability strictly higher than classical optimal ones.}}
        \label{fig_3dRAC}
\end{figure}

Similarly to the derivations in the previous section, we here compute the parameters for $n=3$, where we have $m_x\in \{1,2,3 \}$. The average success probability in a $(3,d)\!\mapsto\! 1$ QRAC protocol is obtained as follows
\begin{equation}
	\overline{\mathrm{P}}_{Q}=\frac{1}{3}\left(1\!+\!\frac{1}{d}\right)\!+\!\frac{\sqrt{d\!+\!8}+\sqrt{4 d^2\!-\!11d\!+\!8}}{6 d^{3/2}},
 \label{qf3d}
\end{equation}
where as the classical optimal one is given by 
\begin{equation}
	\overline{\mathrm{P}}_{cl}=\frac{1}{3}\left\{1\!+\!\frac{3}{d}\!-\!\frac{1}{d^2}\right\}. \label{cf3d}
\end{equation}
The comparison of quantum and classical probabilities above is shown in Fig.~\ref{figa_3dRAC} and their difference, i.e. $\mbox{Adv}(\mathcal{S}_Q)$, is shown in Fig.~\ref{figb_3dRAC}. It turns out that the quantum strategy with two measurement settings achieves a higher probability over a classical optimal one. 

\subsubsection{Case $\mathbf{n=4}$ and $\mathbf{d}\geq 2 $~.--}

\begin{figure}[t!]
     \centering
     \begin{subfigure}[t]{0.42\textwidth}
         \centering
         \includegraphics[scale=0.5]{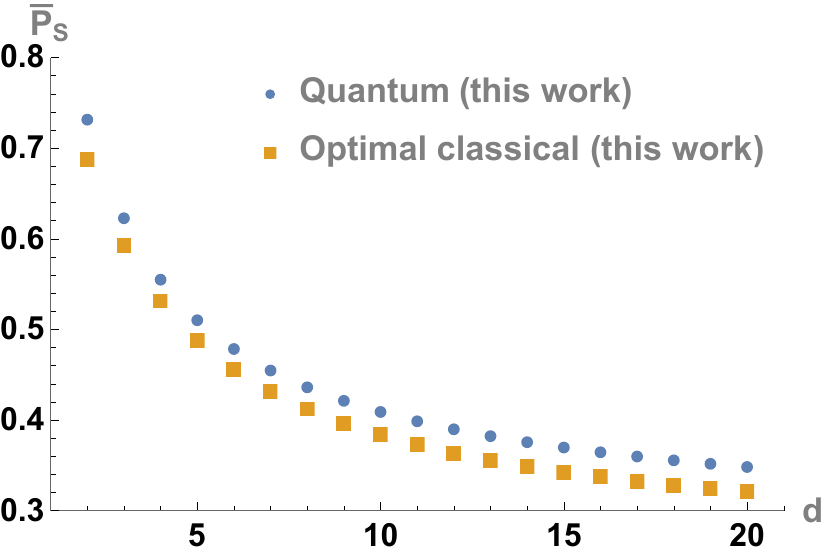}
         \caption{{\small Quantum and optimal classical probabilities in Eqs. (\ref{qf4d}) and (\ref{cf4d}) are shown for $d\leq 20$, in which quantum advantages are achieved.} }
          \vspace{0.5cm}
         \label{figa_4dRAC}
     \end{subfigure}
     \hfill
     \begin{subfigure}[t]{0.42\textwidth}
         \centering
         \includegraphics[scale=0.5]{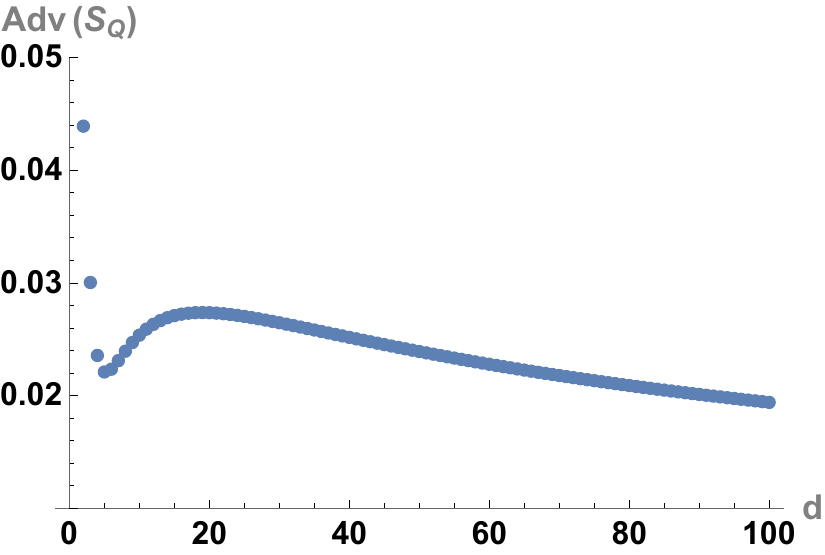}
         \caption{{\small The difference $\mbox{Adv}(\mathcal{S}_Q)$ is computed up to $d\leq 100$.}}
         \label{figb_4dRAC}
     \end{subfigure}
        \caption{ {\small Quantum advantages in $(4,d)\!\mapsto\! 1$ RACs for $d\leq 100$ are shown. The quantum protocol with two MUBs shows a success probability strictly higher than classical optimal ones.}}
        \label{fig_4dRAC}
\end{figure}

For $n=4$, we have $m_x \in \{1,2,3,4\}$. In this case, we have a quantum success probability,
\bea
\overline{\mathrm{P}}_{Q} & = & \frac{1}{4} \left(1+\frac{5}{2d}-\frac{1}{d^2} \right) +\frac{1}{4 d^{5/2}} \left(  \sqrt{ d+3}  + (d-2)\times \right. \nonumber \\
&& \left. \sqrt{ d^2 - 3 d +3 } +\frac{3}{2} \sqrt{ d^3+4 d^2-16 d+12 } \right) ~~~\label{qf4d}
\eea
and, also, we obtain a closed form of classical optimal one for $n=4$ and arbitrary $d$ values,
\begin{equation}
\overline{\mathrm{P}}_{cl}=\frac{1}{4} \left(1+\frac{6}{d}-\frac{7}{d^2}+\frac{4}{d^3}\right).\label{cf4d}
\end{equation}
The comparison of two probabilities are shown in Fig.~(\ref{figa_4dRAC}), and the difference $\mbox{Adv}(\mathcal{S}_Q)$ is computed up to $d\leq 100$, see Fig.~(\ref{figb_4dRAC}).  

\subsection{Computation for $\mathbf{(n,d)\!\mapsto\! 1}$ RAC protocols}

\begin{algorithm}[t]
    \caption{Computing Quantum Value From Our Two Measurement Protocol}
    \textbf{Input:} N $\leftarrow$ upper bound on word length\\
    \textbf{Input:} D $\leftarrow$ upper bound on alphabet size
    
    DEFINE M, initialized to 0, as a two-dimensional array of size $N \times D$\;
    
    \For{n $\leftarrow$ 2 \KwTo N}{
        \For{d $\leftarrow$ 2 \KwTo D}{
            VAL $\leftarrow$ 0\;
            
            \For{k $\leftarrow$ 0 \KwTo $d^n - 1$}{
                x $\leftarrow$ ConvertToBase(k, d, n)\;
                \tcp{ConvertToBase(k, d, n) gives number k in base d with n digits}
                
                $m_x$ $\leftarrow$ Max(LetterCounts(x))\;
                
               $ L_x$ $\leftarrow$ Set of letters with count $m_x$\;
               $ L'_x$$\leftarrow$ $L_x - \{x_n\}$\;  
                $l^{\star}$ $\leftarrow$ FirstElement ($L'_x$)\;  
                
                VAL $\leftarrow$ VAL + $v_Q(x)$ [ by using Eq.~(\ref{eq:Qval})]\;  
            }
            M[n][d] $\leftarrow$ VAL\;
        }
    }
    
    \textbf{Output:} M\;
\end{algorithm}

We have explored the comparison of quantum and classical success probabilities, where the quantum probability is obtained with a two-measurement protocol and the classical probability is proven to be optimal. 


For other cases, up to $n\leq 6$ and $d\leq 6$, quantum probabilities are computed in Table \ref{table2} and their gap with classical optimal probabilities are shown in Table \ref{table3}. In all of the cases, we find that quantum probabilities can be explicitly computed and they are strictly higher than classical optimal ones. 

Our results can be applied to a $(n,d)\mapsto 1$ RAC protocol in general and compute an average quantum success probability. Let us summarize the algorithm to compute quantum probabilities. 


\begin{table}[t!]
\centering
\begin{tabular}[b]{c||ccccc } 
			${\rm n}$~$\backslash$~${\rm d}$ & 2 & 3 & 4 & 5 & 6 \\ [2pt]
			\hline \hline \\
			2~~~~~ & ~$0.85$~ & ~$0.79$~ & ~$0.75$~ & ~$0.72$~ & ~$0.70$~\\ [6pt]
			3~~~~~ & ~$0.77$~ & ~$0.66$~ & ~$0.60$~ & ~$0.56$~ & ~$0.54$~\\ [6pt]
			4~~~~~ & ~$0.73$~ & ~$0.62$~ & ~$0.55$~ & ~$0.51$~ & ~$0.48$~ \\ [6pt]
			5~~~~~ & ~$0.70$~ & ~$0.58$~ &~ $0.52$~ & ~$0.48$~ &~ $0.45$~\\ [6pt]
			6~~~~~ & ~$0.68$~ & ~$0.55$~ &~ $0.49$~ &~ $0.44$~ &~ $0.42$~\\ [6pt]
			\hline 
		\end{tabular}
		\caption{{\small Success probabilities of QRACs are computed.} \label{table2}}
\end{table}

\begin{table}[t!]
\centering
\begin{tabular}[b]{c||ccccc } 
			${\rm n}$~$\backslash$~${\rm d}$ & 2 & 3 & 4 & 5 & 6 \\ [2pt]
			\hline \hline \\
			2~~~~~ & ~$0.10$~ & ~$0.12$~ & ~$0.13$~ & ~$0.12$~ & ~$0.12$~\\ [6pt]
			3~~~~~ & ~$0.020$~ & ~$0.028$~ & ~$0.037$~ & ~$0.042$~ & ~$0.046$~\\ [6pt]
			4~~~~~ & ~$0.044$~ & ~$0.030$~ & ~$0.024$~ & ~$0.022$~ & ~$0.022$~ \\ [6pt]
			5~~~~~ & ~$0.013$~ & ~$0.024$~ &~ $0.022$~ & ~$0.019$~ &~ $0.017$~\\ [6pt]
			6~~~~~ & ~$0.026$~ & ~$0.016$~ &~ $0.016$~ &~ $0.015$~ &~ $0.014$~\\ [6pt]
			\hline 
		\end{tabular}
 \caption{{\small Quantum advantages $\mbox{Adv}(\mathcal{S}_Q)$ in Eq.~(\ref{eq:qa}) are shown.}}\label{table3}
\end{table}


\section{Remark}\label{sec:5}

On applying our algorithms, in principle, one can compute the success probabilities for any given value of $n$ and $d$. However, computation time grows exponentially with increasing value of $d^n$; i.e., the total number of possible $n$ length words formed from $d$ letters in the alphabet set determines the time taken for our algorithm to give the results of computation. Remarkably, quantum advantages in RAC protocols are evidenced for all the cases where we find closed form formulae, as well as for all cases where we applied our algorithm, for example in the range $(n\leq 11,~d\leq 6)$ and $(n\leq 6,~7\leq d\leq 21)$ (see~\ref{Appendix-B}), the quantum strategy achieves a higher success probability than an optimal classical one.

We could not consider instances with other values of $(n,d)$ due to the limitation on computational resources. We conjecture that the quantum advantages would appear in $(n,d) \mapsto 1$ RACs in general and leave it as an open question. The problem also addresses minimal resources, i.e., two measurement settings, to achieve quantum advantages in information theory. 





\begin{figure}[t!]
     \centering
     \begin{subfigure}[h]{0.4\textwidth}
         \centering
         \includegraphics[scale=0.6]{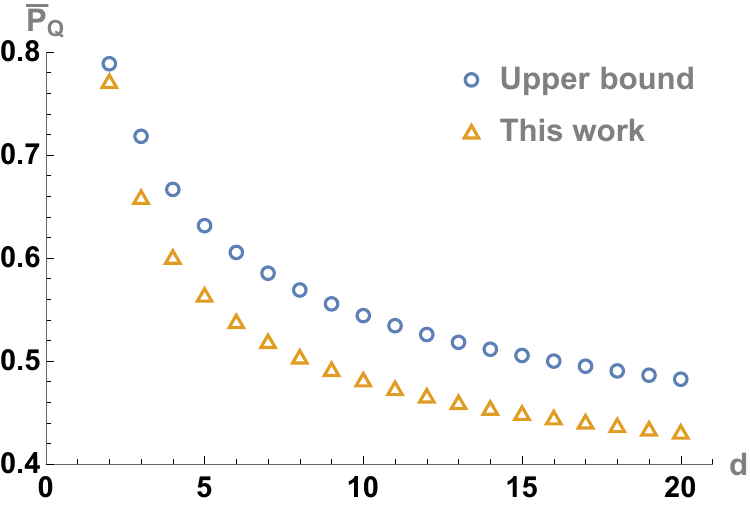}
         \caption{{\small $(3,d)\!\mapsto\! 1$ RACs }}
         \label{fig5a}
     \end{subfigure}
     \hfill
     \vspace{0.5cm}
     \begin{subfigure}[h]{0.4\textwidth}
         \centering
         \includegraphics[scale=0.6]{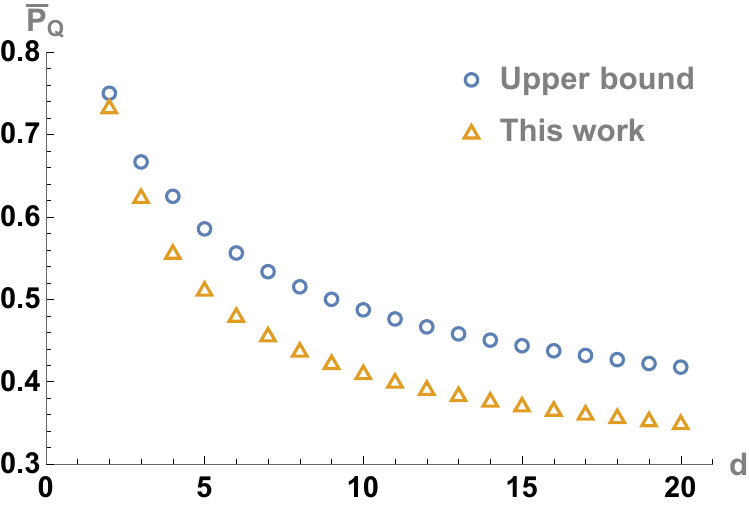}
         \caption{{\small $(4,d)\!\mapsto\! 1$ RACs.}}
         \label{fig5b}
     \end{subfigure}
         \hfill
         \vspace{0.5cm}
     \begin{subfigure}[h]{0.4\textwidth}
         \centering
         \includegraphics[scale=0.6]{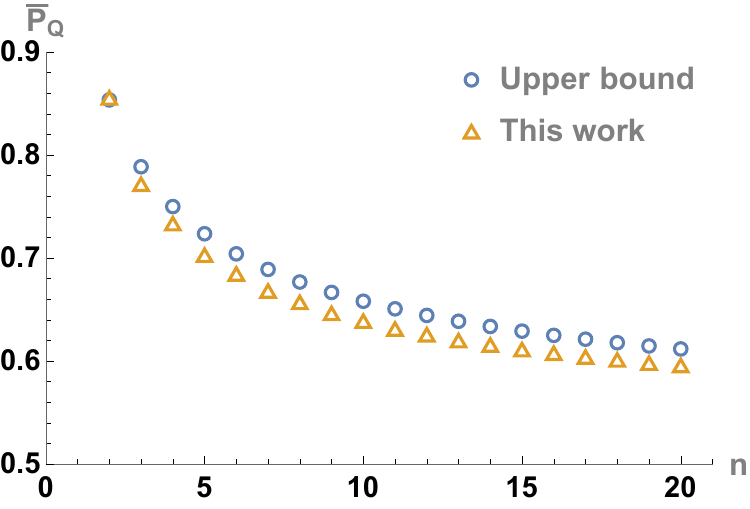}
         \caption{{\small $(n,2)\!\mapsto\! 1$ RACs.}}
         \label{fig5c}
     \end{subfigure}
      \hfill
    \vspace{0.5cm}
        \caption{{\small $(n,d)\!\mapsto\! 1$ RACs.-- The plots depict the average success probability of our quantum protocol and upper bounds on the optimal quantum value derived in Ref.~\cite{FarkasMiklinTavakoli2024}.}}
        \label{fig5}
\end{figure}

Recently, simple and general bounds to QRAC protocols have been reported in Ref. \cite{FarkasMiklinTavakoli2024} and are compared to our results as follows. Firstly, upper bounds are provided,
\begin{align}\label{FMTbound}
    \overline{\mathrm{P}}_{Q}
\leq	\begin{cases}
		\frac{1}{d}\left(1+\frac{d-1}{\sqrt{n}}\right) & \mbox{if}~ n \geq d, \\[8pt]
		\frac{1}{n}\left(1+\frac{n-1}{\sqrt{d}}\right)&\mbox{if}~ n < d.
	\end{cases}  
\end{align}
These bounds are strictly higher than the success probabilities in Eqs. (\ref{qf3d}) and (\ref{qf4d}), see also Fig.~(\ref{fig5}). 

Secondly, three incompatible measurements have been applied in the decoding in \((3,d) \mapsto 1\) QRAC protocols, i.e., $n=3$, and can be compared to cases with two measurements, see Table \ref{table:comp}. It turns out that three measurements can lead to a higher success probability in the protocol. That is, although two incompatible measurements suffice to achieve a success probability higher than classical strategies, up to $n=3$, they would not reach the quantum optimal values. In future investigations, it would be interesting to investigate the tightness of the upper bounds and the usefulness of more than two measurement settings for $n>3$.



\begin{table}[t!]
\begin{center}
\begin{tabular}{|c|c|c|c|c|}
\hline 
\shortstack{~$d$~~\\~}& \shortstack{~\\ $(\overline{\rm P}_{cl})_{max}$\\ (this work)} & \shortstack{~\\ $\overline{\rm P}_Q$\\(this work)} & \shortstack{~\\ $(\overline{\rm P}_Q)_{max}$~\cite{FarkasMiklinTavakoli2024} \\ (lower bound)} & \shortstack{~\\ $(\overline{\rm P}_Q)_{max}$~\cite{FarkasMiklinTavakoli2024} \\ (upper bound)} \\
\hline \hline
~~2~~ & 0.750 & 0.770 & 0.789 & 0.789  \\[3pt] \hline
~~3~~ & 0.630 & 0.657 & 0.697  & 0.718 \\[3pt] \hline
~~4~~ & 0.563 & 0.599 & 0.644  & 0.667 \\[3pt] \hline
~~5~~ & 0.520 & 0.562 & 0.610  & 0.631 \\[3pt] \hline
~~6~~ & 0.491 & 0.536 & 0.582  & 0.605 \\[3pt] \hline
\end{tabular}
\caption{{\small In a $(3,d)\!\mapsto\! 1$ RACs, for $d\in\{2,3,4,5,6\}$, the table lists the values of average success probabilities: (i) optimal classical value, (ii) quantum values from our two measurement protocol, (iii) lower bounds on optimal quantum value from see-saw method, and (iv) upper bounds on optimal quantum value derived in Ref.~\cite{FarkasMiklinTavakoli2024}.}} \label{table:comp}

\end{center}
\end{table}

\section{Conclusion} 
\label{sec:6}

In conclusion, we have presented the complete characterization of optimal classical strategies in \((n,d) \mapsto 1\) RACs and shown how a quantum strategy beats the classical limitations. We have shown that the classical strategy, the MEID protocol in Ref. \cite{Tavakoli}, is optimal in that it achieves the highest classical success probability. Then, we have shown that a single measurement in decoding cannot have a quantum advantage, i.e., a higher success probability than a classical strategy, in a QRAC protocol. 

We have presented a QRAC protocol with the decoding with two incompatible measurements, one from the computational basis and the other from the Fourier basis, and shown that it can go beyond the classical optimal strategies; the quantum advantage is explicitly presented. When applying our protocol to RACs with word lengths \(n=2,3,4\) and any finite alphabet size \(d (>2)\), we derive closed-form formulas for the success probability and prove that a quantum advantage always exists. Such formulas require counting arguments and can, in principle, be derived for other values of \(n\) and arbitrary \(d\), but the complexity of the counting continues to increase. It is very likely the quantum advantage would appear in general for higher values of $(n,d)$, which we leave an open question. 

Our results can be compared to recent results. Firstly, our QRAC protocol with the decoding with two incompatible measurements may be generalized to three measurements, which have been reported recently for $n=3$ with a higher success probability~\cite{FarkasMiklinTavakoli2024}. Hence, we conclude that the decoding with two incompatible measurements can beat the classical bound but is yet sub-optimal. Secondly, the encoding for words are single-qudit quantum states constructed as superposition of two elements from the measurement bases. Contrasting to the quantum protocol in Ref.~\cite{Tavakoli} for $(n\in\{2,3\},d)\rightarrow 1$ RACs, we exploit the freedom to build the encoding states such that amplitudes of the constituent basis states depend on the word to the sender. This provides us with more flexibility in finding quantum advantages by optimizing over a larger class of possible quantum encodings. We find that our approach becomes beneficial when one aims to show quantum advantages in a general $(n,d)\rightarrow 1$ RACs. Then, we find an optimal ansatz by selecting the parameters in the encoding to maximize the quantum gain over the optimal classical value. 

 \section*{Acknowledgements} 

AA and DK acknowledge funding through the Latvian Quantum Initiative under European Union Recovery and Resilience Facility Project No. 2.3.1.1.i.0/1/22/I/CFLA/001. SS acknowledges funding through PASIFIC program call 2 (Agreement No. PAN.BFB.S.BDN.460.022 with the Polish Academy of Sciences). This project has received funding from the European Union’s Horizon 2020 research and innovation programme under the Marie Sk{\l}odowska- Curie grant agreement No 847639 and from the Ministry of Education and Science. JB and AR are supported by the National Research Foundation of Korea (NRF-2021R1A2C2006309, NRF-2022M1A3C2069728), the Institute of Information \& communications Technology Planning \& Evaluation (IITP) grant (Grant No. 2019-0-00831, the ITRC Program/IITP-2022-2018-0-01402).\\

{\em Note added}.-- This manuscript supersedes the preliminary work \cite{arXiv2015AKR} (by some of the authors of the current article), which was unpublished and contained a proof of the optimality of classical strategies. Two later works~ Ref.~\cite{Deba-Pawlowski2018RACs,Deba+PRA2023}  provide a different proof (with similar reasoning) showing that the MEID strategy for classical $(n,d)\!\mapsto\!1$ RACs (or a generalization~\cite{Deba+PRA2023}) is optimal. We like to emphasize here that, our proof is different and novel in that it characterizes all possible optimal classical protocols and it is a proof-by-construction. Our alternative proof provides a full characterization of optimal classical strategies.Since we fully map the optimal protocols through a sequence of non-value-decreasing protocols, we gain some new insights that how the classical optimal protocols are achieved. Also some insights gained from the construction of our proof lead us to find a quantum protocol  by considering a minimal set of measurements for decoding, consisting of only two incompatible measurements. Our quantum protocol is advantageous possibly for all $(n,d)\rightarrow 1$ RACs; a conjecture supported by the computational algorithm and generated results provided in our work.

\appendix

\section{Characterization of all classical optimal strategies}
\label{appendix-A}
In addition to the results proven in the main text, here we provide a complete characterization of all the optimal classical strategies. Let us start first by defining a property that has been stated in the main text.

\begin{pr}\label{property1}
A decoding matrix is said to satisfy property-$1$ if all the elements in each column of the matrix are distinct.
\end{pr}
Also, we recall here the definition of all decoding matrices that satisfy the property-\ref{property1}.
\begin{defn}[Set $F^{\star}$] A set of all decoding matrices satisfying the property-\ref{property1} is defined as $F^{\star}$. If a decoding matrix is in the set $F^{\star}$, we denote it by $f^{\star}$.
			 \label{defFstar}
	\end{defn}
In the following lemma, we will state and prove that for almost all $(n,d)\mapsto 1$ RACs, the optimal classical strategies are fully characterized by the set $F^{\star}$.

 \begin{lem}\label{lemma004}
    For all $(n, d) \mapsto 1$ RACs satisfying either $[d > 2 \text{ and } n > 2]$ or $[d = 2 \text{ and } n \text{ is odd}]$, and for any choice of a column index $j$ and two distinct row indices $y_1$ and $y_2$, the value of any strategy with Property \ref{property1}, i.e., any $f^{\star} \in F^{\star}$, will decrease when the value $f^{\star}_{y_2j}$ is revised to $f^{\star}_{y_1j}$.
\end{lem}

	\begin{proof}[Proof]
		We prove the statement for $y_1=0,~ y_2=1$, and $j=1$, but the same arguments will also hold for any pair of distinct rows $f^{\star}_{y_1\ast},~f^{\star}_{y_2\ast}$ and any column index $j$.
	Property \ref{property1} means that $f^{\star}_{01} \neq f^{\star}_{11}$. We call $f^{\star}_{01}=a$ and $f^{\star}_{11}=b$. We now compare the strategy $f^{\star}$ (where $f^{\star}_{01}=a$ and $f^{\star}_{11}=b$) with the strategy $f$ (where $f_{01}=a$ and $f_{11}=a$). Also note that, to prove that the value decreases on a change of strategy $f^{\star}\!\rightarrow\! f$, equivalently we can show the reverse, i.e., a change of strategy from $f\!\rightarrow\! f^{\star}$ will increase the value. Then, below we will prove that $\mbox{Value}(f^{\star})>\mbox{Value}(f)$.
		
		From the proof of the Lemma \ref{lemma002} it is already known that $\mbox{Value}\left(f^{\star}\right)\ge \mbox{Value}\left(f\right)$ since, in the reference strategy $f$ (i.e., considering the transition from $f$ to $f^{\star}$), there is an injective mapping from the set of "suffering" $a$-words, that is words from the set $A_f$ to the set of "benefiting" $b$-words, i.e, words in the set $W_f^b$. So now it is sufficient to show that there is no inverse mapping, i.e., there are more $b$-beneficiaries than $a$-sufferers. Here we recall, from the sets defined in the proof of Lemma \ref{lemma002}, that the set of all "suffering" $a$-words and  all "benefiting" $b$-words are respectively
		\begin{eqnarray}
			A_f&\!=\!&\{x^a \!=\! a~x_2~...~x_n~\vert~ \mbox{sim}(x,f_{1\ast})\!>\!\mbox{sim}(x,f_{y\ast})~\forall~y \! \neq \!1\}, \nonumber\\
			W^b_f&\!=\!& \{x^b \!=\! b~x_2~...~x_n~\vert~ \mbox{sim}(x,f_{1\ast})\! \geq \! \mbox{sim}(x,f_{y\ast})~\forall~y \! \neq \!1\}.
		\end{eqnarray}
		Now all we need to show is that there is a $b$-word $x=b~x_2~...~x_n$ such that $x^b\in W^b_f$ but the corresponding $a$-word $x^a=a~x_2~...~x_n \notin A_f$. For this purpose we construct $x^b$ as follows, we take a $n$-length word starting with the letter $b$, then the next $k=\lfloor(\frac{n-1}{2})\rfloor$ number of letters are taken from the corresponding positions on the string $f_{1\ast}$, and after that, next $k=\lfloor(\frac{n-1}{2})\rfloor$ number of letters are taken from the corresponding positions on the string $f_{0\ast}$. In case $n$ is even one last position in the word remains to be assigned, in such a situation we assign a letter $c\notin \{f_{0n},~f_{1n}\}$ to the $n^{th}$ position. The word $x^b$ then is as follows
		\begin{eqnarray}
			x^b
			\!=	\! \begin{cases}
				b~f_{12}~...~f_{1(k\!+\!1)}~f_{0(k\!+\!2)}~...~f_{0n}~& \mbox{if}~n~\mbox{is~odd},  \vspace{0.4cm}\\
				b~f_{12}~...~f_{1(k\!+\!1)}~f_{0(k\!+\!2)}~...~f_{0(2k\!+\!1)}~c~& \mbox{otherwise}. 
			\end{cases} 
		\end{eqnarray}
		Now note that for the word $x^b$, we have $\mbox{sim}(x^b,f_{1\ast})=\mbox{sim}(x^b,f_{0\ast})=k$. On the other hand, for all $y\notin\{0,1\}$ we get $\mbox{sim}(x^b,f_{y\ast})=0$ because the first column of $f$ does not have letter $b$ and in each of the remaining columns letters do not repeat. Therefore, $x^b \in W^b_f$. However, the corresponding $a$-word $x^a$ (by replacing the first letter in $x^b$ by $a$) satisfies $\mbox{sim}(x^a,f_{1\ast})=\mbox{sim}(x^a,f_{0\ast})=k+1$, and for all $y\notin\{0,1\}$ one gets $\mbox{sim}(x^a,f_{y\ast})=1$. This implies that $x^a \notin A_f$, which proves our claim.
		Finally, we note that our construction and proof hold for all RACs with either $[~d>2~~\&~~n>2~]$, or $[~d=2~~ \&~~n~\mbox{is~odd}~]$.
	\end{proof}
	
	We just showed that whenever, either $[~d>2~~\&~~n>2~]$, or $[~d=2~~ \&~~n~\mbox{is~odd}~]$, any strategy which does not satisfy Property \ref{property1}, can be gradually improved to a strictly better one with Property \ref{property1}, and that no better result can be achieved. Next, we characterize all the optimal decoding strategies for the two exceptions: (i) the cases where $d=2$ and $n$ is even, and (ii) the cases where $n=2$. Let us define another property to cover the remaining cases.

 \begin{pr} \label{property2}
 A decoding matrix $f$ is said to satisfy property-$2$ if, except for exactly one column, all elements in the other columns are distinct. 
	\end{pr}

 \begin{defn}[Set $F^{\circ}$] The set of all decoding strategies which satisfy the property-\ref{property2} is defined as $F^{\circ}$. If a decoding matrix is in the set $F^{\circ}$, we denote it by $f^{\circ}$.
			 \label{defFcirc}
	\end{defn}
\textbf{Notations:} Note that $F^{\circ} \cap F^{\star} = \varnothing$. The column of $f^{\circ} \in F^{\circ}$ with repeated entries is denoted by $j^{\circ}$. If a decoding matrix is in either the set $F^{\star}$ or the set $F^{\circ}$, we denote it by $f^{\ostar}$; that is, $f^{\ostar} \in F^{\circ} \cap F^{\star}$.

	\begin{lem}\label{lemma005}
			For all $(n,2)\!\mapsto\! 1$ RACs where $n(>2)$ is even, all decoding strategies $f^{\ostar}\in F^{\circ} \cap F^{\star}$, i.e., those satisfying either Property \ref{property1} or Property \ref{property2}, have the same value. Moreover, for all decoding strategy $f^{\circ}\in F^{\circ}$, for any two row indices $y_1$ and $y_2$ and any column $j\neq j^{\circ}$, the value of the strategy will decrease on changing $f^{\circ}_{y_2j}$ to $f^{\circ}_{y_1j}$. 
	\end{lem}
	
	\begin{proof}[Proof] 
We first show that for all $f^{\circ}\in F^{\circ}$, any change of strategy $f^{\circ}\!\rightarrow\! f$ such that $f \notin F^{\circ} \cap F^{\star}$ leads to decrease in value. We note that the proof will run very similar to the proof of Lemma \ref{lemma004}. Without loss of generality, we make a change to the $j=1$ column of $f^{\circ}$ assuming that the two elements in this column are distinct, suppose the $n^{th}$ column is the one with both elements the same, i.e., $j^{\circ}=n$. Say $f^{\circ}_{01}=a$ and $f^{\circ}_{11}=b$, then change $f^{\circ}\!\rightarrow\! f$ is such that $f=f^{\circ}$ except, at exactly one position, $f_{11}=a$. Now, equivalently, we will show the opposite, that a change $f^{\circ}\!\rightarrow\! f$ will increase the value. Similar to the argument provided in proof of the Lemma \ref{lemma004}, we can construct a word
		\begin{eqnarray}
			x^b&=&b~f_{12}~...~f_{1(k\!+\!1)}~f_{0(k\!+\!2)}~...~f_{0(2k\!+\!1)}~c,\\ &~&\mbox{where}~~ c\neq f_{0n}=f_{1n}. \nonumber
		\end{eqnarray}
		Then, it follows that the word $x^b \in W^b_f$ (the set of "benefiting" $b$-words) but the corresponding word $x^a \notin A$ (the set of "suffering" $a$-words). Therefore, $\mbox{Value}(f^{\circ})>\mbox{Value}(f)$.
		
		Next, we show that for all $f^{\circ}\in F^{\circ}$, any change of strategy $f^{\circ}\!\rightarrow\! f^{\star}$ keeps the value of the strategy same, i.e., $\mbox{Value}(f^{\star})=\mbox{Value}(f^{\circ})$. Note that for all strategies $f^{\circ}$ such a transformation is always possible in one step. Without loss of generality, we consider that the first column of $f^{\circ}$ has both elements the same, say $a$. Then, let's say we change $f^{\circ}\!\rightarrow\! f^{\star}$ by replacing the second element of the firsts column by $b\neq a$, i.e., in the new strategy $f^{\star}_{11}=b$ (at all other entries $f^{\star}=f^{\circ}$). We recall that, with respect to the strategy $f^{\circ}$, the set of all "suffering" $a$-words and  all "benefiting" $b$-words are respectively
		\begin{eqnarray}
			A_{f^{\circ}}&=&\{x^a=a~x_2~...~x_n~\vert~ \mbox{sim}(x,f^{\circ}_{1\ast})>\mbox{sim}(x,f^{\circ}_{0\ast})\}, \nonumber\\
			W^b_{f^{\circ}}&=& \{x^b=b~x_2~...~x_n~\vert~ \mbox{sim}(x,f^{\circ}_{1\ast})\geq \mbox{sim}(x,f^{\circ}_{0\ast})\}. 
		\end{eqnarray}
		Since $n$ is even, in fact, for the considered case definition of the set of all "benefiting" $b$-words reduces to, 
		\begin{eqnarray}
			W^b_{f^{\circ}}= \{x^b=b~x_2~...~x_n~\vert~ \mbox{sim}(x,f^{\circ}_{1\ast})>\mbox{sim}(x,f^{\circ}_{0\ast})\}, 
		\end{eqnarray}
		because if a $b$-word match with $f^{\circ}_{1\ast}$ at even number of position it must match with $f^{\circ}_{0\ast}$ at odd number of positions, and the two number-of-matching-positions must add to $n-1$ (an odd number) and vise versa. Now we can see that, there is a one-to-one and onto correspondence between the set of words in $A_{f^{\circ}}$ and $W^b_{f^{\circ}}$. Therefore, $\mbox{Value}(f^{\star})=\mbox{Value}(f^{\circ})$.
	\end{proof}
		\begin{figure}[t!]
		\begin{center}
			\includegraphics[scale=0.3]{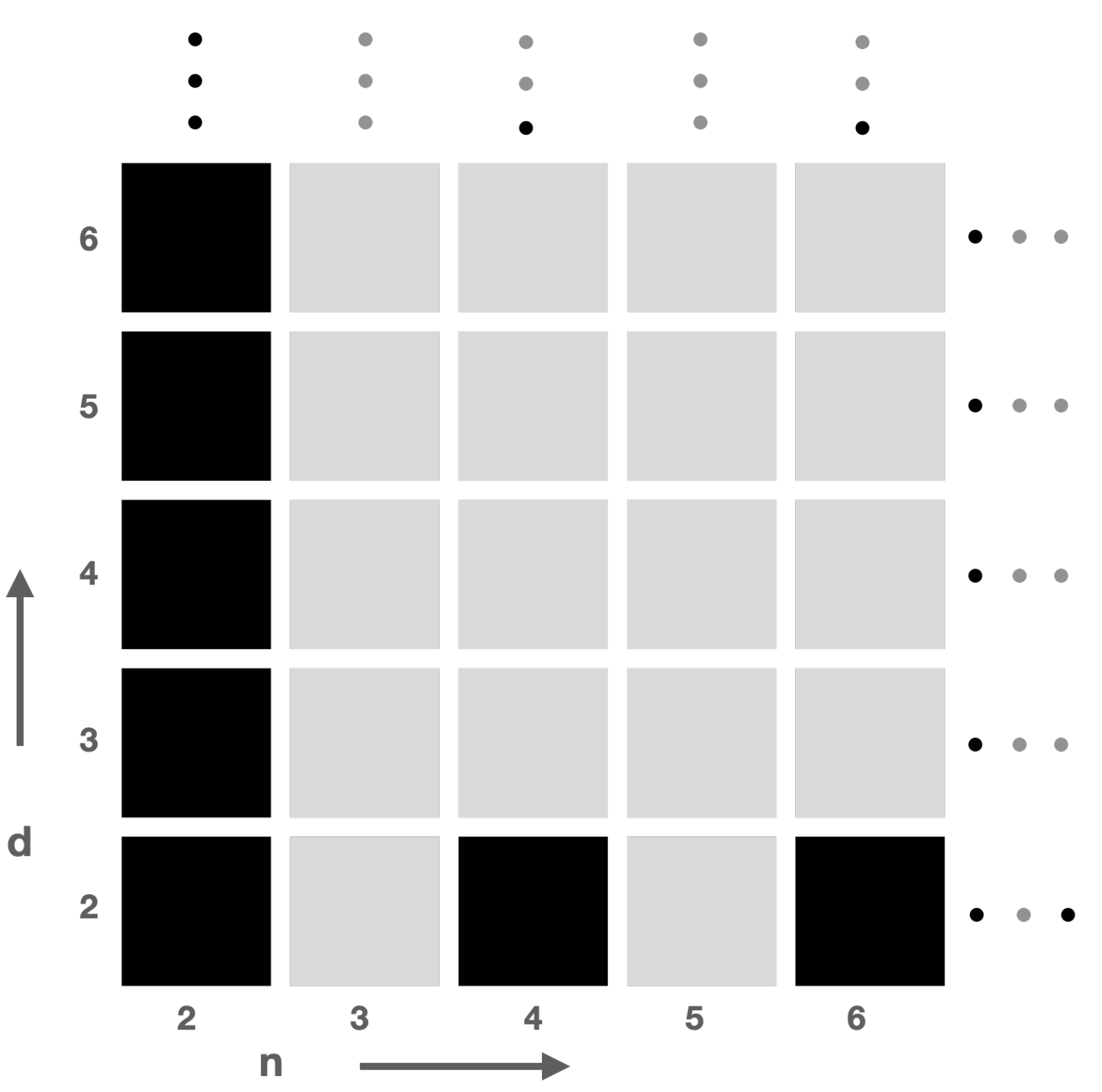}
			\caption{{\small The figure shows that: (i) for values of $(n,d)$ at the (light) grey cell positions, Property~\ref{property1} characterizes all possible optimal decodings, and (ii) for values of $(n,d)$ at (dark) black cell positions decoding functions satisfying either Property~\ref{property1} or Property~\ref{property2} characterizes all possible optimal strategies.}}
			\label{Figure2}
		\end{center}
	\end{figure}
	\begin{lem}\label{lemma006}
			For all $(2,d)\!\mapsto\! 1$ RAC, all decoding strategies $f^{\ostar}\in F^{\star} \cup F^{\circ}$, i.e., those satisfying either Property \ref{property1} or Property \ref{property2}, have the same value. Moreover, for all decoding strategy $f^{\circ}\in F^{\circ}$, for any two row indices $y_1$ and $y_2$ and any column $j\neq j^{\circ}$, the value of the strategy will decrease on changing $f^{\circ}_{y_2j}$ to $f^{\circ}_{y_1j}$. 
	\end{lem}
	
	\begin{proof}[Proof]
		 We first show that the value of all decoding strategies $f^{\ostar}\in F^{\circ} \cap F^{\star}$ is the same. We consider that all elements in the first column of $f^{\ostar}$ are distinct, and its second column can be filled arbitrarily. The proof will run similarly on exchanging the property of the first with the second column. There are total $d^2$ possible words $x=x_1x_2\in X^2$. Consider the following subset of words: $X^2_{f^{\ostar}}=\{f^{\ostar}_{y\ast}~\vert~ y\in\{0,...,d\!\!-\!\!1\}\}$. Now, $x\in X^2_{f^{\ostar}}\!\Rightarrow\! \mbox{max}_y ~{\rm sim}(x,f^{\ostar}_{y\ast})=2$, and $x\notin X^2_{f^{\ostar}}\!\Rightarrow\! \mbox{max}_y ~{\rm sim}(x,f^{\ostar}_{y\ast})=1$ (since all entries of first column of $f^{\ostar}$ are distinct). Therefore, 
		\begin{eqnarray}
			v_{f^{\ostar}}(x)
			=	\begin{cases}
				\frac{1}{d^2} ~~~~& \mbox{if}~ x\in X^2_{f^{\ostar}},\\
				\frac{1}{2 d^2} ~~~~& \mbox{otherwise}. 
			\end{cases}  
		\end{eqnarray}
		So, $\mbox{Value}(f^{\ostar})=\frac{d+1}{2d}$ for all $f^{\ostar}\in F^{\circ} \cap F^{\star}$. 
		
		Next, suppose we consider another strategy $f$ revised from a given one $f^{\circ}$ where the first column of $f$ consists of a repeated entry. That is, $f$ is identical to $f^{\circ}$ except one entry where $f_{y_{2}1}=f^{\circ}_{y_{1}1}=a$~ for $y_1\neq y_2$. To obtain $f$, an element $f^{\circ}_{y_21}=b$ is replaced by $a$. Then, we have $x\in X^2_{f}=\{f_{y\ast}~\vert~ y\in\{0,...,d\!\!-\!\!1\}\}\!\Rightarrow\! \mbox{max}_y ~{\rm sim}(x,f_{y\ast})=2$ ~(~and $d\!-\!1\leq |X^2_{f}|\leq d$~). On the other hand, $x\notin X^2_{f}\!\Rightarrow\! \mbox{max}_y ~{\rm sim}(x,f_{y\ast})\leq 1$. However, if $d>2$, there is at least one word $x^b=bc$, where $c\notin \{f_{y_12},f_{y_22}\}$, such that $x^b\notin X^2_{f}$ and $v_f(x^b)=0$, therefore $\mbox{Value}(f)<\mbox{Value}(f^{\circ})$. If $d=2$, for a strategy $f^{\circ}$ both the entries in the second column must take the same value, in that case, the change $f^{\circ}\!\rightarrow\! f$ will make both the rows of $f$ identical, so from lemma-1, the value will decrease.
  \end{proof}
  
	We just showed in Lemma \ref{lemma005} and Lemma \ref{lemma006} that in two special cases, along with all the decoding strategies satisfying Property \ref{property1} ($F^{\star}$), all those satisfying Property \ref{property2}  ($F^{\circ}$) are also optimal strategies. In all other (almost all) cases, all optimal strategies are exactly those satisfying Property \ref{property1} ($F^{\star}$). All the exceptional cases covered by Lemma \ref{lemma005} and Lemma \ref{lemma006} are depicted as dark-shaded squares in Figure~(\ref{Figure2}). So we conclude our characterization of all optimal (deterministic) strategies as in the following theorem.

 \begin{thm}\label{theorem001c}
    For an $(n,d) \mapsto 1$ RAC, when either $[d > 2 \text{ and } n > 2]$ or $[d = 2 \text{ and } n \text{ is odd}]$, all optimal decoding matrices are exactly those for which all entries in any given column are distinct, i.e., all members of the set $F^{\star}$. In the remaining two cases (analyzed in detail in Lemmas \ref{lemma005} and \ref{lemma006}), all decoding matrices in $F^{\star}$ are optimal, and there are additional optimal decoding matrices with the property that, except for exactly one column, all elements in any other column are distinct, i.e., all members of the set $F^{\circ}$.
\end{thm}

\section{Quantum advantage for $(n,d)\!\mapsto\! 1$ RACs}
\label{Appendix-B}
In the following two tables, Table~\ref{tabB1} and Table~\ref{tabB2}, we list the quantum advantage from our protocol computed by applying our algorithm.
{\small \begin{table}[h!]
\centering
\begin{tabular}[b]{c||ccccc } 
			${\rm n}$~$\backslash$~${\rm d}$ & 2 & 3 & 4 & 5 & 6 \\ [2pt]
			\hline \hline \\
			2~~~~~ & 0.103553 & 0.122008 & 0.125 & 0.123607 & 0.120791 \\[3pt]
 3~~~~~ &~~~0.0180651~~~ & ~~~0.0201777~~~ & ~~~0.0307403~~~ & ~~~0.0375291~~~ & ~~~0.0417375~~~ \\[3pt]
 4~~~~~ &0.0399016 & 0.0283296 & 0.0169907 & 0.0151703 & 0.0157141 \\[3pt]
 5~~~~~ &0.0112877 & 0.0199864 & 0.0201164 & 0.0147433 & 0.0116019 \\[3pt]
 6~~~~~ &0.021557 & 0.0101198 & 0.0124627 & 0.0132385 & 0.0112675 \\[3pt]
 7~~~~~ &0.00738842 & 0.0103394 & 0.00643974 & 0.0074583 & 0.00820718 \\[3pt]
 8~~~~~ &0.0135951 & 0.00825719 & 0.0060108 & 0.00431117 & 0.00447647 \\[3pt]
 9~~~~~ &0.00511049 & 0.00465557 & 0.00569363 & 0.00373259 & 0.00271772 \\[3pt]
 10~~~~~ &0.00936507 & 0.00505481 & 0.00354345 & 0.00347664 & 0.00231769 \\[3pt]
 11~~~~~ &0.00368065 & 0.00405002 & 0.00230845 & 0.00256693 & 0.0021063 \\[3pt]
		\end{tabular}
 \caption{{\small Quantum advantage for $(n,d)\!\mapsto\! 1$ RACs: $2\leq n \leq 11$ and $2\leq d \leq 6$.}}\label{tabB1}
 \vspace{0.4cm}
\end{table}}


{\small \begin{table}[h!]
\centering
\begin{tabular}[b]{c||ccccc } 
			${\rm d}$~$\backslash$~${\rm n}$ & 2 & 3 & 4 & 5 & 6 \\ [2pt]
			\hline \hline \\			
7~~~~~ & ~~~0.117554~~~ & ~~~0.0443487~~~ & ~~~0.0168861~~~ & ~~~0.00996495~~~ & ~~~0.00923988~~~ \\[3pt]
8~~~~~ & 0.114277 & 0.0459561 & 0.0181291 & 0.0092355 & 0.00761972 \\[3pt]
9~~~~~ & 0.111111 & 0.0469147 & 0.0192632 & 0.00902587 & 0.00645251 \\[3pt]
10~~~~~ & 0.108114 & 0.0474429 & 0.0202407 & 0.00910577 & 0.00566282 \\[3pt]
11~~~~~ & 0.105301 & 0.0476792 & 0.021061 & 0.00934105 & 0.00515991 \\[3pt]
12~~~~~ & 0.102671 & 0.0477142 & 0.0217394 & 0.00965369 & 0.00486641 \\[3pt]
13~~~~~ & 0.100214 & 0.0476087 & 0.0222954 & 0.00999829 & 0.0047227 \\[3pt]
14~~~~~ & 0.0979163 & 0.0474046 & 0.0227476 & 0.0103486 & 0.00468468 \\[3pt]
15~~~~~ & 0.0957661 & 0.0471311 & 0.0231125 & 0.0106897 & 0.00472029 \\[3pt]
16~~~~~ & 0.09375 & 0.0468089 & 0.0234045 & 0.0110135 & 0.00480639 \\[3pt]
17~~~~~ & 0.091856 & 0.0464529 & 0.0236353 & 0.0113159 & 0.00492631 \\[3pt]
18~~~~~ & 0.0900734 & 0.0460742 & 0.0238148 & 0.0115953 & 0.00506807 \\[3pt]
19~~~~~ & 0.0883921 & 0.0456808 & 0.0239514 & 0.0118514 & 0.00522306 \\[3pt]
20~~~~~ & 0.0868034 & 0.0452785 & 0.0240517 & 0.0120849 & 0.00538511 \\[3pt]
21~~~~~ & 0.0852994 & 0.0448718 & 0.0241215 & 0.0122969 & 0.00554981 \\[3pt]
		\end{tabular} 
 \caption{{\small Quantum advantage for $(n,d)\!\mapsto\! 1$ RACs: $7\leq d \leq 21$ and $2\leq n \leq 6$.}}
 \label{tabB2}
\end{table}}

\section*{References} 
\vspace{1cm}
\bibliographystyle{iopart-num} 
\bibliography{qracs} 

\end{document}